# Translocatome: a novel resource for the analysis of protein translocation between cellular organelles

Péter Mendik[1], Levente Dobronyi[1], Ferenc Hári[1], Csaba Kerepesi[2,3], Leonardo Maia-Moço[1,4], Donát Buszlai[1], Peter Csermely[1] * and Daniel V. Veres[1,5]

[1]Department of Medical Chemistry, Semmelweis University, Budapest, Hungary;
[2]Institute for Computer Science and Control (MTA SZTAKI), Hungarian Academy of Sciences, Budapest, Hungary;
[3]Institute of Mathematics, Eötvös Loránd University, Budapest, Hungary;
[4]Cancer Biology and Epigenetics Group, Research Center of Portuguese Oncology Institute of Porto, Portugal
[5]Turbine Ltd., Budapest, Hungary;

* To whom correspondence should be addressed. Tel: +36-1-459-1500 extension: 60130; Fax: +36-1-266-3802; Email: csermely.peter@med.semmelweis-univ.hu

**ABSTRACT.** Here we present Translocatome, the first dedicated database of human translocating proteins (URL: http://translocatome.linkgroup.hu). The core of the Translocatome database is the manually curated data set of 213 human translocating proteins listing the source of their experimental validation, several details of their translocation mechanism, their local compartmentalized interactome, as well as their involvement in signalling pathways and disease development. In addition, using the well-established and widely used gradient boosting machine learning tool, XGBoost, Translocatome provides translocation probability values for 13,066 human proteins identifying 1133 and 3268 high- and low-confidence translocating proteins, respectively. The database has user-friendly search options with a UniProt autocomplete quick search and advanced search for proteins filtered by their localization, UniProt identifiers, translocation likelihood or data complexity. Download options of search results, manually curated and predicted translocating protein sets are available on its website. The update of the database is helped by its manual curation framework and connection to the previously published ComPPI compartmentalized protein-protein interaction database. As shown by the application examples of merlin (NF2) and tumor protein 63 (TP63) Translocatome allows a better comprehension of protein translocation as a systems biology phenomenon and can be used as a discovery-tool in the protein translocation field.

**INTRODUCTION**

Subcellular localization of proteins is essential in spatial and temporal organisation of biological processes such as signalling pathways enabling their separation into organelles (1). Translocating proteins play a key role in the reconfiguration of cellular functions after environmental changes, as well as in embryonic or disease development. Different subcellular organelles have well characterized interactomes (2,3). With the advance of imaging techniques subcellular dynamics became a rapidly expanding research area (4,5). Restoring or affecting the cellular localization of disease-related proteins emerges as an efficient therapeutic method (6,7).

Protein translocation is a process which refers to the alteration of a given protein's subcellular localization. However, this phenomenon has no unified definition, and the word "translocation" may also refer to gene translocation or RNA translocation at the ribosome. In this work we define protein translocation as a systems biology phenomenon, which refers to the regulated movement of a protein of a given post-translational state between subcellular compartments. Translocation changes the interaction partners and leads to altered function(s) of translocating proteins. There are certain processes (such as co-translational, post-translational delivery-type, cell division-induced, downregulation- or passive diffusion-related phenomena; for their detailed description see Supplementary Texts S1 and S2) that may change the localization of a protein, but to increase the focus and clarity of our database we did not consider them as translocation.

There are widely used protein databases that contain information on protein translocation, e.g. the MoonProt (8) or UniProt (9) databases. However, these databases are not dedicated collections of translocating proteins. Here we present Translocatome, which is a manually curated database of 213 human translocating proteins with extensive information on their translocation. Moreover, Translocatome contains 13 066 human proteins with predicted likelihood of translocation. With the help of the well-established and widely used gradient boosting machine learning tool, XGBoost (10-12) we predicted 1133 high-confidence translocating proteins. In addition, Translocatome contains 3268 and 8665 low-confidence and non-translocating proteins, respectively. To train the XGBoost algorithm, we also created a manually curated set of 139 non-translocating proteins as part of the database. Translocatome is a novel, dedicated database of human translocating proteins including their interaction partners in the different subcellular localizations. This database contributes to a better understanding of protein translocation as a systems biology phenomenon and facilitates further discoveries of translocating proteins. As translocating proteins are already targeted pharmaceutically (6,7) new findings in this field may also lead to better therapeutic options.

## DESCRIPTION OF THE DATABASE

### Overview of Translocatome

Translocatome is the first database that collects manually curated human translocating proteins including their interacting partners in the localizations involved, translocation mechanism (including protein structure details if available), type of experimental evidence, affected signalling pathway(s) and pathological properties. The core of the Translocatome database is the 213 manually curated human translocating proteins (http://translocatome.linkgroup.hu/coredata) which were all collected based on related publications containing experimental evidence. Altogether Translocatome contains 13 066 human proteins, which were selected from the compartmentalized protein-protein interaction database (3; ComPPI http://comppi.linkgroup.hu, downloaded on 20/07/2018) using the inclusion criterion that every protein needed to have at least one experimentally validated subcellular localization. By the application of the well-established gradient boosting machine learning tool, XGBoost (10-12) we predicted 1133 high-confidence translocating proteins. All the 13 066 human proteins were characterized by their translocation likelihood named as Translocation Evidence Score (TES) calculated by the XGBoost machine learning algorithm (Figure 1). Various search and download options make it possible for users to process these data according to their goals.

### Database content

The core data of Translocatome is the extensively curated set of 213 human translocating proteins (see Core Data at the website: http://translocatome.linkgroup.hu/coredata). With the manual curation process involving the judgement of 3 independent experts we aimed to collect detailed and experimentally validated information about every entry extracted from peer reviewed publications (for the details of the manual curation process see Supplementary Text S3, Supplementary Table S1 and Supplementary Figure S1). For each of the 213 manually curated translocating proteins we collected the available subset of the following data:

- a.) name set, gene name and UniProt (9) accession number and link,
- b.) PubMed ID(s) and link(s) to peer-reviewed article(s) describing the experimental evidence of translocation,
- c.) initial and target localizations of the translocating protein,
- d.) interacting partners and biological functions (both in the initial and target compartments),
- e.) translocation mechanism,
- f.) the used detection method,
- g.) protein structural information on translocation mechanism,
- h.) disease group, exact disease involved and pathological role,
- i.) signalling pathways affected.

We used the UniProt naming convention (9) for protein identification, Gene Ontology terms (13,14) for localization/biological process identification and the KEGG naming convention (15) for the standardization of signalling pathways. Following the logic of our previously published compartmentalized protein-protein interaction database (ComPPI, 3) every protein was annotated with one of six major cellular localizations (cytoplasm, extracellular space, mitochondria, nucleus, membrane or secretory-pathway). If there was more precise localization information available it was included as a minor localization. All 213 manually curated translocating proteins are characterized by a Data Complexity Score (DCS) as described later in detail, which makes it easier to assess the amount of information associated with each protein. 53 of the manually curated proteins showed translocation exclusively under pathological conditions (such as cancer). Therefore, we used the remaining 160 physiologically translocating proteins as a positive training set (Supplementary Table S2) for the widely used XGBoost machine learning algorithm (10-12).

We also collected a manually curated negative dataset of 139 human non-translocating proteins, each one classified as a protein a.) with experimentally proved diffuse, multi-compartmental distribution, b.) with exclusive single-compartment localization, c.) docked to DNA/RNA, d.) embedded in membranes or e.) attached to the cytoskeleton (for additional details see Supplementary Text S4). These 139 proteins were used as a negative training set (Supplementary Table S3) for the application of the XGBoost machine learning algorithm (10-12). For a detailed description of our database structure see Figures 1A and 1B.

Altogether Translocatome contains 13 066 human proteins having at least one experimentally validated localization as described in our in house developed compartmentalized protein-protein interaction database (ComPPI, 3). From the ComPPI database we also imported the interactome of these human proteins having 151 889 interactions. The translocation likelihood of all the 13 066 proteins is characterized by a Translocation Evidence Score (TES) as described later in detail. The translocation likelihood was calculated by the XGBoost machine learning algorithm (10-12) as detailed in the next Section.

**The XGBoost machine learning algorithm-based prediction of translocating proteins**

The machine learning procedure followed the general methodology of supervised machine learning workflow: data collection, feature extraction, feature selection, classification, training, testing and interpretation. For each step we applied an existing, well-characterized approach. Data collection and feature extraction were based on established procedures as described below. For all additional steps we applied the well-established, widely used gradient boosting-type (10) machine learning tool, XGBoost (11). XGBoost was successfully applied in hundreds of recent studies to predict e.g. host-pathogen protein-protein interactions (16), microRNA disease association (17) and DNA methylation (18). Several studies including our own previous paper showed that XGBoost gives the best performance if compared with a number of known machine learning methods (see e.g. Refs. 12, 16 and 18).

To train the XGBoost method first we annotated each of the 13 066 proteins of the Translocatome database with their relevant Gene Ontology (GO, 13,14) cellular component, biological process and molecular function terms also including their ancestors. This resulted in 21 020 annotated GO terms total (all details of the methodology are available here: https://github.com/kerepesi/translocatome_ml). The process was based on our previous work (12), for its details please see Supplementary Text S5.

Next, each of the 13 066 proteins were annotated with their degree and bridgeness in the compartmentalized protein-protein interaction database (ComPPI, 3) derived human interactome containing 151 889 interactions. Degree (the number of human interactome neighbours) was included, since the 213 manually curated translocating proteins showed a significantly higher degree than that of the 139 manually curated non-translocating proteins or the average (Supplementary Figure S2). This is not surprising since translocating proteins often have a central role in regulation behaving as interactome hubs. Similarly, translocating proteins often connect interactome modules (large protein mega-complexes), thus act as bridges. Indeed, the 213 manually curated translocating proteins had significantly higher bridgeness values than that of the 139 manually curated non-translocating proteins or the average (Supplementary Figure S2). Degree and bridgeness values were calculated using the CytoScape network analyser program (19) and its ModuLand plug-in (20), respectively. GO terms, degree and bridgeness formed the feature sets selected by the XGBoost machine learning method.

Since the human interactome (3) we used for the calculation of degree and bridgeness did not contain interactions observed in pathological conditions, we excluded those 53 of the manually curated proteins from the positive training set of the XGBoost algorithm, which showed translocation exclusively under pathological conditions (such as cancer). The remaining 160 manually curated proteins were used as the positive training set (Supplementary Table S2).

Following the methodology of several XGBoost studies (11,16-18) including our previously published work (12) we evaluated the XGBoost-selected feature sets by 5-fold cross-validation, and we evaluated their predictive power by the area under the curve of the receiver operating characteristic curve (ROC AUC or shortly AUC, 21). 5-fold cross-validation is a widely used method where the training data is split into five random parts and four parts are used to train the XGBoost machine learning tool and the prediction of the fifth part is evaluated. For every feature set, we repeated this process 100 times. We selected those GO features which had a feature important value (produced by the XGBoost program) greater than 0.02. With this generally applied XGBoost procedure we reached an average AUC of 0.916 (±0.0046 standard deviation) with only 15 GO features left from the initial 21 020 (see Table 1). We continued feature selection by adding the two interactome-derived features degree and bridgeness using the giant component of the ComPPI-derived human protein-protein interaction network (1). In these calculations the giant component of the interactome was used which did not contain 9 proteins of the total. The inclusion of the two network-related features produced an average AUC of 0.9207 (±0.0056 standard deviation), showing a further increase from the average AUC of 0.916 and implying a high performance. We show the ROC curves of 100 five-fold cross-validation runs of the final feature set on Figure 1C having a minimal, average and maximal AUC of 0.9047, 0.9207 and 0.9333, respectively. As shown on Supplementary Figure S3 both precision-recall and Matthews correlation coefficient curves also showed a high performance of the learning process. For more details of the generally applied machine learning procedure see Supplementary Text S6. All data of the procedure are available at https://github.com/kerepesi/translocatome_ml along with codes to reproduce the results.

The feature set of the XGBoost model with the best AUC value is shown on Table 1. Features with positive importance values increase the probability of translocation. These are Gene Ontology features from each main GO category (cellular components, biological processes and molecular functions), which are often associated with protein translocation as described in Table 1 in detail. If a feature has a negative importance value, then it decreases the probability of translocation. Two categories of low degree and low bridgeness values each, as well as 6 GO-terms negatively associated with protein translocation are listed among these negative features. Using the feature set shown on Table 1 we calculated the Translocation Evidence Score characterizing the translocation probability of each of the 13 066 proteins in the Translocatome database as described in the next section.

# Data Complexity and Translocation Evidence Scores

*Data Complexity Score.* To provide an easy assessment of the information available of a manually curated protein we developed the Data Complexity Score (DCS). DCS varies between 0 and 1, having increasing values if the protein has more curated data. The score is calculated and normalized after weighting all the available data, where those related to translocation have a higher weight (please find the detailed calculation process in Supplementary Text S7). Therefore, DCS is not only shows the quantity but also the relevance of the available data. In addition DCS indicates which entries may require further curation.

*Translocation Evidence Score.* The XGBoost machine learning method gave every protein of the Translocatome database a Translocation Evidence Score (TES) that is proportional with the translocation probability of the given protein. For each protein we computed TES using Equation 1

$$\sum_{i=1}^{n} w_i x_i \qquad (1),$$

where $w_i$ is the importance value of the *i* th feature of the model (see *i* th row of Table 1). The importance value was calculated as described in the legend of Table 1. $x = 1$, if the given feature is true for that protein and $x = 0$, if it is false (*n* is the number of features of the model; here *n=19*). TES values were rescaled to the interval [0,1] by min-max normalization using Equation 2

$$x' = \frac{X - X_{min}}{X_{max} - X_{min}}. \qquad (2),$$

The larger the TES value, the greater the probability of translocation. As a numerical example, suppose that "protein A" has 20 neighbours (degree) in the human interactome and its UniProt record contains only two GO terms, "animal organ morphogenesis", and "cytoplasm". Then the predicted translocation evidence score of "protein A" is *-0.497 + 2.675 + 1.353 = 3.531*. The value is then normalized using Equation 2. For each of the 13 066 proteins, the respective TES scores can be found both in the search results and in the downloadable datasets.

The Translocation Evidence Score gave the possibility to define a cut-off value, below which proteins were considered as non-translocating. To define this cut-off value, we used the widely used measure of a test's accuracy, the F1 score (also called as F-measure, 22) that measures the performance of a binary classification being a harmonic average of precision and recall (also called as sensitivity). Supplementary Figure S4 shows recall, fallout, precision and the F1 score at different threshold values and illustrates the distribution of the TES values. The F1 score reached its maximum at the threshold of -0.02958, which gives a straightforward cut-off value for translocation probability. Thus proteins having lower TES values than 0.4487 were considered as non-translocating (for more details see Supplementary Texts S6 and S8). In order to give an assessment of potential false positive predictions we also defined a higher TES cut-off value separating low- and high-confidence translocating proteins. We set this value as 0.6167, since above this threshold there were not any negative set proteins. We assume that the probability of false positive predictions is low above this threshold value. Low-confidence translocating proteins, which have a translocation evidence score (TES) between the two threshold values are presumably translocating but they need further validation.

The two Translocation Evidence Score cut-off values separated our original 13 066 human proteins to 3 classes: a.) 1133 high-confidence translocating proteins having a TES value higher than 0.6167; b.) 3268 low-confidence translocating proteins having a TES value between 0.6167 and 0.4487, as well as c.) the residual 8665 proteins having a TES value lower than 0.4487, which were considered as non-translocating (Figure 1B).

**Search, download options and output**

As part of the user-friendly interface, various search functions were developed. We provide an easy to use quick search function (with UniProt AC autocompletion) which can be used to find protein families or a given protein. The advanced search option creates the possibility to search for more elaborate sets of proteins filtered by their localization, UniProt identifiers, Translocation Evidence Score or Data Complexity Score. The web interface provides eight pre-defined protein sets as download options covering 1.) 213 manually curated translocating proteins, 2.) 160 physiologically translocating manually curated proteins (the positive training set), 3.) manually curated non-translocating proteins (the negative training set), 4.), 5.) and 6.) high-, low-confidence and non-translocating protein sets, as well as 7.) the whole protein set and 8.) its protein-protein interaction network. These sets of proteins can be downloaded in a comma separated .csv format. Besides these pre-defined sets users can also download the results of their search queries as a tabulator separated file (.tsv, see the technical parameters in the "Design and implementation" section). Examples and explanations of the output formats are available in Supplementary Figures S5, S6 and S7.

**Design and implementation**

To allow the development of the Translocatome database as a community effort a manual curation framework (MCF) was designed. MCF uses the same MongoDB database as the Translocatome site, with a user interface developed in the Ruby on Rails 4.2 (https://rubyonrails.org) framework. The MCF website follows the hierarchical model-view-controller design pattern to ensure the separation of the data layer from the business logic and the user interface. The MCF stores all the data of the Translocatome and provides them to the front-end of the Translocatome website after expert review. To ease usability an end-user documentation is available as tutorials, detailed descriptions and location-specific tooltips in the HELP menu on the site (http://translocatome.linkgroup.hu/help). Further details of design and implementation of the database are summarized in Supplementary Text S9.

**Application examples**

The Translocatome database is the only dedicated collection of human translocating proteins. With its Translocation Evidence Score (TES) for 13 066 proteins it helps the identification and experimental validation of novel translocating proteins. To demonstrate the prediction efficiency of Translocatome we assessed the first 40 proteins with the highest TES values. Table 2. shows the list of the best performing 25 proteins. They fall into 4 categories: A.) were already included in the manually curated 213 translocating protein set (12 proteins: PTEN, PTK2, FOXO3, GMNN, ATF2, MAPK1, GLI3, HRAS, AR, SMAD3, SMAD2 and HSP90AB1); B.) were previously shown to be translocating proteins but have not appeared in our Core Data of 213 proteins collected from keyword-based searches (11 proteins: NF2, TULP3, SNCA, FGFR2, MTOR, GSK3B, EIF6, HDAC1, CARM1, CUL1 and RARB; see Supplementary Table S4); C.) have not been described as translocating proteins yet, but from the literature we can conclude that their translocation is probable (1 protein: TP63); D.) there is no information in the literature about their translocation (1 protein: PRKRA). Proteins of categories c.) and d.) are good candidates for further experimental studies verifying their translocation.

The best hit of the XGBoost algorithm, the PTEN protein is a part of the manually curated 213 translocating proteins. As its second best hit, the XGBoost algorithm correctly predicted NF2 (Merlin) as a translocating protein, since NF2 in its dephosphorylated form indeed translocates to the nucleus (23). NF2 is a hub having 48 neighbours and was characterized by 6 out of the 15 Gene Ontology terms that were important according to the best XGBoost model predicting translocation.

Out of the 25 proteins listed on Table 2, the p63 protein (tumor protein 63, TP63) is the only protein, which falls into the category C.) containing "proteins having implications in the literature that they are translocating". p63 is not tagged as translocating in available databases (8,9). p63 is a protein that is physiologically found in the nucleus of human cells (Figure 2). It acts as a transcription factor either activating or repressing specific DNA sequences (24) and it is an essential factor during embryogenesis (25). Besides these conventional functions it is also known that p63 appears in the cytoplasm of adenocarcinoma or prostate carcinoma cells. Moreover, the cytoplasmic localization of p63 results in the increased malignancy of these tumours (26,27). This disease-altered localization of p63 is in compliance with our definition for a translocating protein. Thus, the XGBoost machine learning algorithm correctly predicted the translocation of p63. As p63 is associated with poor survival of cancer patients (26,27) its targeting may serve as a therapeutic option.

With the above examples we demonstrated that the XGBoost machine learning algorithm (10-12) is able to classify previously known proteins effectively and may also predict new translocations correctly. Out of the 25 best hits shown on Table 2 the PRKRA protein (Interferon-inducible double-stranded RNA-dependent protein kinase activator A) is the only one, which appears to be a completely new translocating protein candidate. It will be an interesting question of further experimental studies, whether this protein is indeed translocating or shuttling between the cytosol and the nucleus as predicted by the rather equal number of its protein interactions (3) in these two compartments.

**Comparison with similar tools**

The existing MoonProt (8) and UniProt (9) databases contain potentially translocating proteins performing multiple biochemical functions or data related to protein translocation, respectively. Out of the **75** human proteins of the latest, 2.0 version of the MoonProt database (accessed on 04/01/2018) **55** proteins were shown in the literature to translocate in a regulated manner (and were included to the Translocatome). The other 20 human moonlighting proteins achieve their multiple functions in the same cellular compartment. Out of the total number of **20 239** human UniProt proteins (accessed on 17/11/2017), we can presume a translocation in **1013** cases based on their UniProt description or subcellular location data. As only **75** (**35%**) of the **213** Translocatome gold standard proteins were included in the **1013** presumably translocating UniProt proteins, the Translocatome database can greatly supplement this aspect of the UniProt database. From the residual **938** UniProt translocation candidates **25**% and **34**% were predicted in the Translocatome as high- and low-confidence translocating proteins, respectively. **31**% of the 938 UniProt proteins was predicted as non-translocating while **10**% of them was not part of the Translocatome database.

**CONCLUSIONS AND FUTURE DIRECTIONS**

In summary, Translocatome offers a unique dataset of 213 specifically collected human translocating proteins listing the source of their experimental validation, several details of their translocation mechanism, local compartmentalized interactome as well as their involvement in signalling pathways and disease development. In addition, it provides translocation likelihood values (as Translocation Evidence Scores) for 13 066 human proteins identifying 1133 and 3268 high- and low-confidence translocating proteins, respectively. The assembly of the Translocatome database (Figure 1) combines careful manual curation steps with a state-of-art machine learning prediction protocol. The application examples (Table 2 and Figure 2) show that the Translocation Evidence Score of Translocatome is able to highlight already experimentally verified translocating proteins, which do not evidently appear by key word-based search methods, as well as proteins, whose translocation is already very likely from the literature, but has not been directly verified yet. These features position Translocatome as a discovery-tool in the field of protein translocation.

The Translocatome database can be accessed via a user-friendly web-interface providing a quick search function (with UniProt AC autocompletion) and an advanced search to find sets of proteins filtered by their localization, UniProt identifiers, Translocation Evidence Score or Data Complexity Score. The web interface provides eight pre-defined protein sets as download options and a possibility to download the search results. End-user documentation is available as tutorials, detailed descriptions and location-specific tooltips in the HELP menu of the site.

Translocatome is available at http://translocatome.linkgroup.hu. Translocatome is a community-annotation resource, which is helped by its manual curation framework (MCF). MCF allows the users to build in their own experimentally verified translocating proteins. Translocatome will be updated and upgraded annually for minimum 5 years. The Translocatome database is connected to our previously developed, compartmentalized protein-protein interaction database (ComPPI, 3). Thus the improvement of the subcellular localization and interactome data can be easily translated to regular updates of the Translocatome database giving improved protein translocation likelihood values.

We plan to resolve current Translocatome limitations, such as extending the database to other species than humans. Future plans include the extension of positive and negative datasets and localization-based network visualization. Translocating RNAs play a key role in subcellular regulation as well, but their role is even more complex and mysterious. We plan to extend our database and add translocating RNAs, to fill out this gap. The improvement of the data not only means, that Translocatome will have more proteins or more detailed information. In this process the whole database will be updated meaning that the XGBoost machine learning will reappraise the data and provide more even accurate predictions based on the updated data.

In conclusion, the Translocatome database introduced here provides the first dedicated collection of 213 translocating human proteins including their interaction partners in the different subcellular localizations. Importantly, Translocatome gives a Translocation Evidence Score to more than 13 thousand human proteins allowing the assessment of their translocation probability. All these features are accessible in a user-friendly manner. The Translocatome database allows a better comprehension of protein translocation as a systems biology phenomenon, and can be used as a discovery-tool of the field. Since translocating proteins become more and more important therapeutic targets (6,7) Translocatome may contribute to the development of better future therapeutic options.

**AVAILABILITY**

The Translocatome database of human translocating proteins can be accessed freely at [http://translocatome.linkgroup.hu](http://translocatome.linkgroup.hu).

**SUPPLEMENTARY DATA**

Supplementary Data are available at NAR online.

**FUNDING**

This work was supported by the Hungarian National Research Development and Innovation Office (OTKA K115378), the New National Excellence Program of the Hungarian Ministry of Human Capacities (to PM and DVV) and by the Higher Education Institutional Excellence Programme of the Ministry of Human Capacities in Hungary, within the framework of molecular biology thematic programmes of the Semmelweis University. CK was supported by the European Union, co-financed by the European Social Fund (EFOP-3.6.3-VEKOP-16-2017-00002) and by the Higher Education Institutional Excellence Programme of the Ministry of Human Capacities in Hungary "2018-1.2.1-NKP-00008: Exploring the Mathematical Foundations of Artificial Intelligence" grant. Publication charge was supported by EFOP-3.6.3-VEKOP-16-2017-00009.

**CONFLICT OF INTEREST**

DVV and PC are founders, DVV is an employee of the startup firm Turbine Ltd (http://turbine.ai).

**ACKNOWLEDGEMENT**

Authors acknowledge the contribution of LINK-Group (http://linkgroup.hu) members: László Mérő in the development of the MCF, as well as Dániel Ferenc Felföldi, Dániel Kisvárday-Papp and Virág Lakner in the collection of manually curated data.

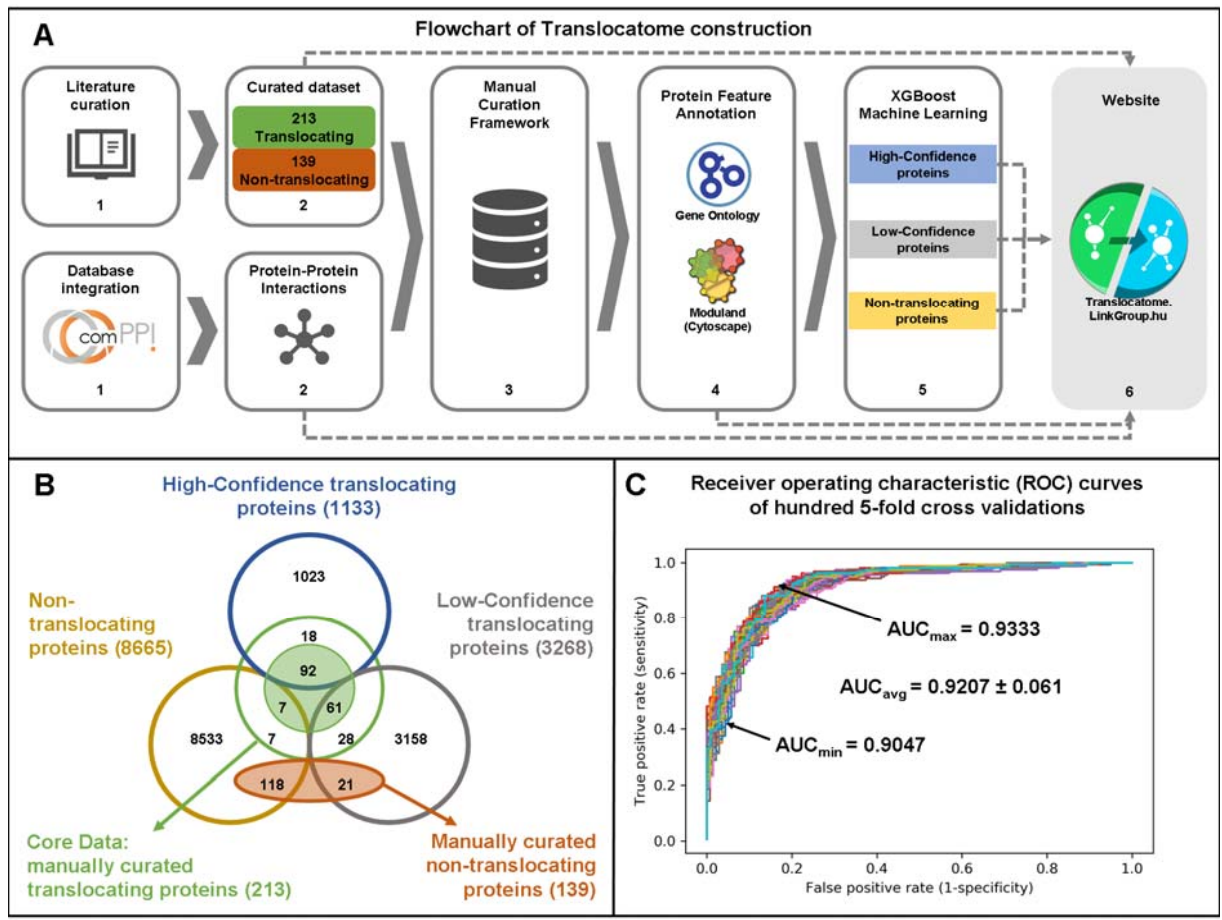

Figure 1. **The structure of the Translocatome database and performance of the XGBoost machine learning prediction method. (A)** Schematic flowchart of the Translocatome database construction process highlighting 6 major steps**.** The panel shows the main input sources of the Translocatome are manual curation of peer reviewed articles and the ComPPI database (http://comppi.linkgroup.hu; 3). In the manual curation process we recorded the source of experimental validation, several details of translocation mechanism, the local compartmentalized interactome, as well as the involvement in signalling pathways and disease development **(1)**. This extensive manual curation resulted in a set of 213 translocating and another set of 139 non-translocating human proteins. To incorporate our data into a Protein-Protein Interaction (PPI) network we imported the PPI of 13 066 ComPPI (3) human proteins with their 151 889 interactions **(2)**. The Manual Curation Framework (MCF) is a user-friendly interface where the data of the Translocatome database is stored and after registration users from all over the world can log in to modify and update its data, which is published as part of the Translocatome database after expert cross-check **(3)**. To enable the prediction of translocating proteins we annotated each protein in our database with Gene Ontology (13,14) functional and ComPPI-derived interactome (3) topological properties **(4)**. The XGBoost machine learning algorithm (10-12) classified the 13 066 human proteins into three sets: high-, low-confidence translocating proteins and non-translocating proteins **(5)**. On the http://translocatome.linkgroup.hu website the whole dataset is available for searching and downloading purposes freely and without registration. Translocatome can be updated by the community-based Manual Curation Framework. Moreover, Translocatome is linked to the ComPPI database (3) so in the case of its update Translocatome can be also updated **(6)**. **(B)** Structure of the Translocatome database. As shown by a Venn-diagram the database consists of the Core Data of 213 manually curated translocating proteins (available here: [http://translocatome.linkgroup.hu/coredata](http://translocatome.linkgroup.hu/coredata)), which are extended by 1133 and 3268 high- and low-confidence translocating proteins, respectively. Green and red filled circles represent the positive and negative training sets, respectively. Core Data and positive learning set differ, since the latter does not contain the 53 proteins showing translocation exclusively under pathological conditions (such as cancer). **(C)** Performance of the widely-used XGBoost machine learning method (10-12) on the final feature set. Each of the 100 different receiver operating characteristic (ROC) curves belong to a different 5 fold cross-validation run on the training set (containing 160 physiologically translocating and 139 non-translocating proteins). In the runs the XGBoost machine learning method used the final feature set (see Table 1) selected earlier as described in the main text and Supplementary Text S6. The minimal, maximal and average area under the curve (AUC) were 0.9047, 0.9333 and 0.9207 (±0.0061 standard deviation), respectively.

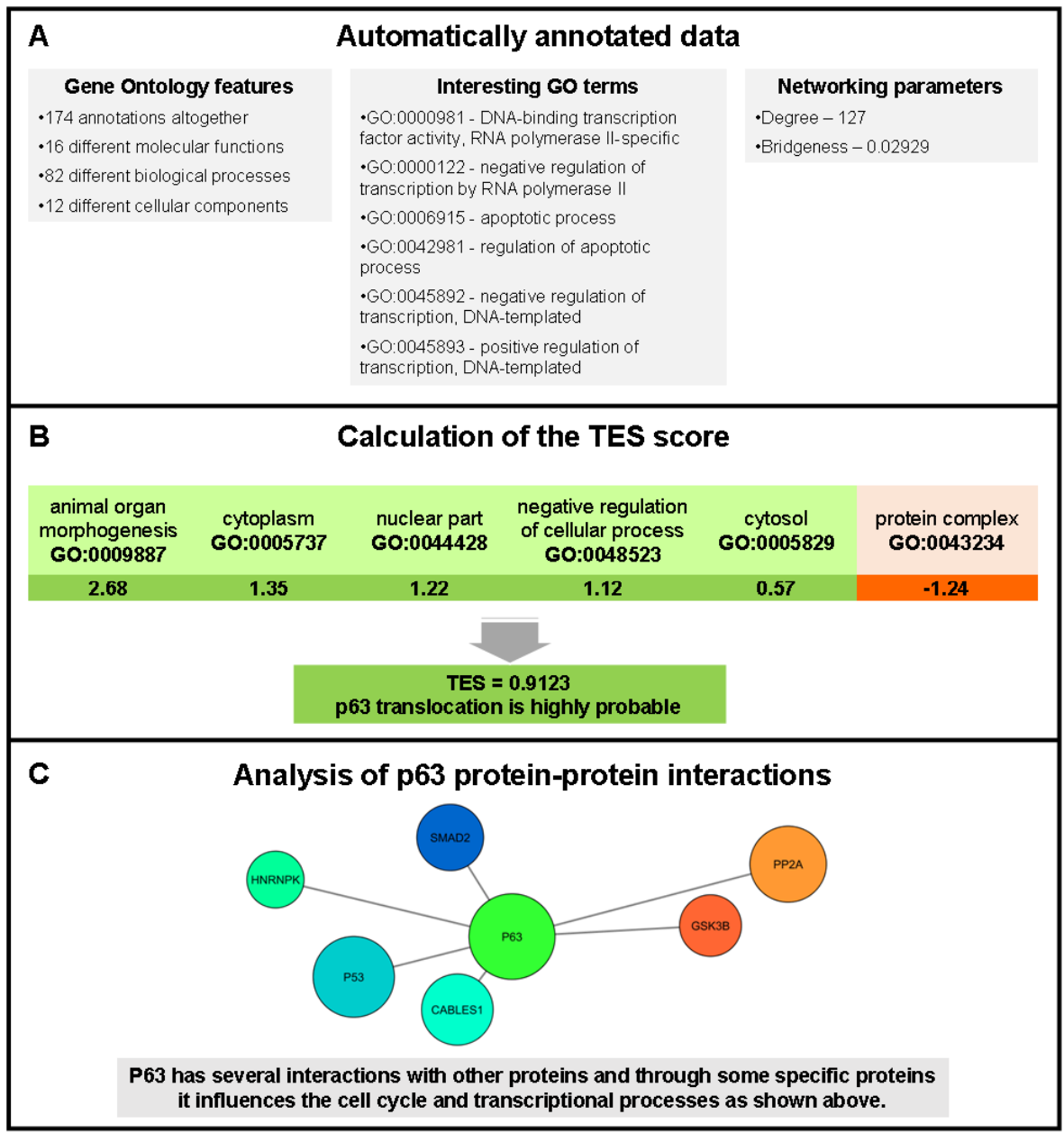

Figure 2. **p63, a translocating protein predicted by the XGBoost machine learning algorithm. (A)** In the left column the Gene Ontology (GO) terms (13,14) that are associated with the p63 protein are summarized, showing that altogether the protein is characterized by 174 annotations. As some of these annotations are redundant, altogether there are 12, 16 and 82 specific GO terms of cellular components, molecular functions and biological processes, respectively. In the right column the degree and the bridgeness value of p63 in the ComPPI database-derived human interactome (3) are shown. In the centre 6 highlighted GO terms show that p63 plays an important role in the regulation of transcription and the apoptotic process. A complete list of associated GO terms was collected by Quick-GO (28) and is available here:
https://www.ebi.ac.uk/QuickGO/annotations?geneProductId=Q9H3D4. **(B)** The XGBoost machine learning algorithm (10-12) selected 17 types of features as the best model when calculating the Translocation Evidence Score (TES, see Table 1 and Supplementary Text S6). Out of the 17 features the p63 protein is characterized by 6 GO features and a large degree. For every GO-related feature we have shown the name of the specific GO term and the respective importance value of this GO term. The high TES score shows that the translocation of p63 is highly probable. **(C)** As it is also suggested by some of its major interaction partners shown, the p63 protein is a regulator of transcription and apoptosis. Reviewing the literature, we found that besides the well-known nuclear localization of the p63 protein (24,25) in fact, it also has a validated cytoplasmic localization, too. Moreover, cytoplasmic localization of p63 is a predictor of increased malignancy of some tumours (26,27). This disease-altered localization of p63 is in compliance with our definition for a translocating protein. Thus p63 was correctly predicted by Translocatome as a likely candidate of further experimental studies proving its translocation.

**Table 1. The feature set identified as best predictor by the XGBoost machine learning algorithm**

| Gene Ontology process (GO term name) or interactome feature | Importance | Short biological explanation |
|---|---|---|
| **Parameters having a positive predictive value** | | |
| *animal organ morphogenesis (GO:0009887)* | 2.68 | Morphogenesis and other developmental processes are mostly regulated through complex networks of transcription factors, where translocation is often involved as a regulation step (29). |
| *regulation of carbohydrate metabolic process (GO:0006109)* | 1.53 | Quite some metabolic enzymes also function as protein kinases and translocate between cellular compartments playing a role e.g. in carcinogenesis (30). |
| *cytoplasm (GO:0005737)* | 1.35 | Large cellular compartments are often associated with proteins that translocate. Nucleo-cytoplasmic translocations play a key role in the regulation of transcription factors (29). |
| *nuclear part (GO:0044428)* | 1.12 | |
| *negative regulation of cellular process (GO:0048523)* | 1.12 | Negative regulatory mechanisms are frequently exerted by translocating proteins such as e.g. PTEN (31) or transcription factors. |
| *plasma membrane part (GO:0044459)* | 0.70 | Large cellular compartments are often associated with proteins that translocate. Cytosol-membrane translocations play a key role in the regulation of signalling pathways (32). |
| *extracellular region (GO:0005576)* | 0.65 | |
| *cytosol (GO:0005829)* | 0.57 | |
| *spliceosomal complex (GO:0005681)* | 0.23 | The spliceosome is constituted by snRNPs translocating from the cytoplasm. Some spliceosome components are also involved in mRNA export (33,34). |
| **Parameters having a negative predictive value** | | |
| *bridgeness value is lower than 0.000292 (bridgeness lower than 0.000292)* | -0.36 | Translocating proteins often bridge the two interactome modules (large protein complexes) of their two localizations. Therefore, their bridgeness values tend to be high (20 and Supplementary Figure S1). |
| *degree is smaller than 62.5 (degree lower than 62.5)* | -0.50 | A reasonably high number of interaction partners often indicates a role in regulation and signal transduction. Many of these proteins are "date-hubs", which may undergo a translocation process. Nevertheless, too many partners could be a characteristics of a multi compartmental housekeeping protein (35 and Supplementary Figure S1). |
| *degree is smaller than 14.5 (degree lower than 14.5)* | -0.54 | |
| *negative regulation of intracellular signal transduction (GO:1902532)* | -0.61 | If the translocation process becomes inhibited, it may often prevent signal transduction. Inhibition often occurs via sequestration by large protein complexes which usually have only one localization (36). |
| *myeloid cell differentiation (GO:0030099)* | -0.74 | Cell adhesion and membrane bound proteins play an important role in myleoid cell differentiation (37,38). Both protein categories are typically non-translocating proteins, which may over-compensate the role of translocating transcription factors in this process. |
| *intrinsic component of membrane (GO:0031224)* | -0.82 | Intrinsic membrane components predominantly do not translocate to other major localizations. |
| *system process (GO:0003008)* | -0.91 | A wide variety of proteins exert their system level biological functions (e.g. secretion of molecules) in a non-translocating manner: cell membrane channels, actin, myosin, etc. |
| *single organismal cell-cell adhesion (GO:0016337)* | -1.06 | Cell adhesion proteins usually have a strictly limited location in the plasma membrane |
| *bridgeness value is lower than 2.5e-06 (bridgness lower than 2.5e-06)* | -1.10 | Translocating proteins often bridge the two interactome modules (large protein complexes) of their two localizations. Therefore, their bridgeness values tend to be high (20 and Supplementary Figure S1). |
| *protein complex (GO:0043234)* | -1.24 | Proteins often fulfil their roles in large protein mega-complexes. These complexes may assist for other proteins to translocate, but their own components do not translocate. |

Features selected by the XGBoost machine learning algorithm (10-12) can be human protein-protein interaction network-related (3) or GO term-related (13,14), as listed in the first column. XGBoost assigns each feature with an importance score (as shown in the third column) which was calculated as the leaf-scores of the one-depth trees of the best XGBoost model. In the fourth column there is a short (and most of the time, very partial) explanation to explain why these features may become selected by the XGBoost machine learning process as best predictors of protein translocation including some key references supporting the explanations.

## Table 2. List of the first 25 proteins having the highest Translocation Evidence Score

| UniProt ID | Gene names | Protein names | Translocation evidence score | Group | Summary |
|---|---|---|---|---|---|
| P60484 | PTEN | Phosphatidylinositol 3,4,5-trisphosphate 3-phosphatase and dual-specificity protein phosphatase | 1.0000 | A | PTEN translocates to the nucleus from the cytoplasm in response to oxidative stress |
| P35240 | NF2 | Merlin | 0.9807 | B | Dephosphorylated merlin translocates to the nucleus (23) |
| O75386 | TULP3 | Tubby-related protein 3 | 0.9802 | B | Membrane association with PIP2 anchors Tub to sequester TULP3 from transport to the nucleus. It also translocates from the plasma membrane to the nucleus upon activation of guanine nucleotide-binding protein G(q) subunit alpha (39). |
| Q05397 | PTK2 | Focal adhesion kinase 1 (FADK 1) | 0.9798 | A | Retinoid acid induced nuclear FAK translocation leads to a reduced cellular adhesion |
| P37840 | SNCA | Alpha-synuclein | 0.9743 | B | Mitochondrial translocation occurs rapidly under as a result of pH changes during oxidative or metabolic stress (40) |
| O43524 | FOXO3 | Forkhead box protein O3 | 0.9740 | A | Dephosphorylated cytoplasmic Foxo is unidirectionally translocated out of the cytoplasm by the nuclear localization signal and Ran GTPase driven nuclear import system. |
| O75496 | GMNN | Geminin | 0.9740 | A | Geminin is excluded from the nucleus during part of the G1 phase and at the transition from G0 to G1. |
| P21802 | FGFR2 | Fibroblast growth factor receptor 2 (FGFR-2) | 0.9703 | B | Under PGF(2alpha) stimulation, FGF-2 and FGFR2 proteins accumulate near the nuclear envelope and co-localize in the nucleus of Py1a cells (41). |
| P15336 | ATF2 | Cyclic AMP-dependent transcription factor ATF-2 | 0.9677 | A | Some drugs as paclitaxel or vemurafenib are inducers of ATF-2 translocation. |
| P42345 | MTOR | Serine/threonine-protein kinase mTOR | 0.9635 | B | Long-term treatment with rapamycin triggers dephosphorylation and cytoplasmic translocation of nuclear rictor and sin1 accompanied by inhibition of mTORC2 assembly (42). |
| P28482 | MAPK1 | Mitogen-activated protein kinase 1 | 0.9480 | A | MAPK1 (ERK2) translocates to the nucleus and mitochondria. |
| P49841 | GSK3B | Glycogen synthase kinase-3 beta | 0.9462 | B | GSK3 translocated to the plasma membrane, along with AXIN, upon Wnt stimulation (43). |
| O75569 | PRKRA | Interferon-inducible double-stranded RNA-dependent protein kinase activator A | 0.9439 | D | There is no information in the literature about the translocation of this protein |
| P10071 | GLI3 | Transcriptional activator GLI3 | 0.9439 | A | Translocates after interaction with ZIC1. |
| P01112 | HRAS | GTPase Hras | 0.9224 | A | Several pathological conditions such as exogenous hyperoxia induce Ras translocation from cytosol to the membrane. |
| P56537 | EIF6 | Eukaryotic translation initiation factor 6 | 0.9208 | B | Increase in intracellular concentration of calcium leads to rapid translocation of eIF6 from the cytoplasm to the nucleus, an event that can be blocked by specific calcineurin inhibitors, such as cyclosporin A (44). |
| P10275 | AR | Androgen receptor (Dihydrotestosterone receptor) | 0.9123 | A | Translocation happens after ligand binding and is mediated by filamin, which is thought to disrupt the association between Hsp90 and the receptor in the cytoplasm. |
| P84022 | SMAD3 | Mothers against decapentaplegic homolog 3 | 0.9123 | A | Activated TGF-beta receptor phosphorylates Smad2 and Smad3, which then form a complex with Smad4 and translocate to the nucleus. |
| Q13547 | HDAC1 | Histone deacetylase 1 | 0.9123 | B | In neuroblastoma cells translocation of HDAC1 was reported to the cytoplasm in response to HSV-1 viral infection (45). |
| Q15796 | SMAD2 | Mothers against decapentaplegic homolog 2 | 0.9123 | A | After phosphorylation of receptor-regulated SMADs (SMAD1, SMAD2, SMAD3, SMAD5 and SMAD8) they are recognized by SMAD 4. This complex translocates to the nucleus. |
| Q86X55 | CARM1 | Histone-arginine methyltransferase CARM1 | 0.9123 | B | Nucleus --> cytosol translocation mainly occurs during mitosis, but it also occurs out of the cell cycle (46). |
| Q9H3D4 | TP63 | Tumor protein 63 (p63) | 0.9123 | C | Nuclear localization of p63 was correlated with nuclear accumulation of p53, whereas the presence of nuclear p63 had no apparent effect on patient survival (24-27). The mechanism remains to be elucidated. |
| P08238 | HSP90AB1 | Heat shock protein HSP 90-beta | 0.9053 | A | Hsp90 has been found in the extracellular region, and also in the nucleus. |
| Q13616 | CUL1 | Cullin-1 | 0.9034 | B | ROC1 promotes CUL1 nuclear accumulation to facilitate its NEDD8 modification (47). |
| P10826 | RARB | Retinoic acid receptor beta | 0.8967 | B | This is a nucleocytoplasmic shuttling protein, AFP may inhibit translocation of RAR-beta into the nucleus via competitive binding to RAR-beta with ATRA (48). |

Every protein are shown in the table with their 3 indicators (UniProt ID, Gene name and Protein name) and Translocation Evidence Score (TES) as defined in the main text. The higher the TES score the higher the probability of translocation. Proteins fall into four categories as shown in the fifth column. A.) The protein was included in the manually curated 213 translocating protein set. B.) The protein did not appear in our keyword-based searches but was previously shown to be a translocating protein. C.) The protein has not been described as a translocating protein yet, but from the literature we can conclude that its translocation is probable (p63 protein, for more information, see Figure 2). D.) There is no information in the literature about the translocation of this protein (PRKRA). Categories C and D are good candidates for further evaluation. Short summary gives a brief description of the translocation mechanism of each protein having a representative publication cited in categories B and C (for references describing the translocation of proteins in category A see the Translocatome database entry of the respective protein).

# Supplementary Material to

# Translocatome: a novel resource for the analysis of protein translocation between cellular organelles

Péter Mendik[1], Levente Dobronyi[1], Ferenc Hári[1], Csaba Kerepesi[2,3], Leonardo Maia-Moço[1,4], Donát Buszlai[1], Peter Csermely[1] * and Daniel V. Veres[1,5]

[1]Department of Medical Chemistry, Semmelweis University, Budapest, Hungary;
[2]Institute for Computer Science and Control (MTA SZTAKI), Hungarian Academy of Sciences, Budapest, Hungary;
[3]Institute of Mathematics, Eötvös Loránd University, Budapest, Hungary;
[4]Cancer Biology and Epigenetics Group, Research Center of Portuguese Oncology Institute of Porto, Portugal
[5]Turbine Ltd., Budapest, Hungary;
* To whom correspondence should be addressed. Tel: +36-1-459-1500 extension: 60130; Fax: +36-1-266-3802;
Email: csermely.peter@med.semmelweis-univ.hu

## Table of Contents

**Table of Contents** (continued from the previous page)

# Supplementary Texts

## Supplementary Text S1. Cellular processes not assessed as translocation

To increase the focus and clarity of the Translocatome dataset we did not consider co-translational; post-translational delivery-type; cell division-induced; downregulation- or passive diffusion-related phenomena as protein translocation despite of the fact that the subcellular localization of a given protein may change in these situations. Please find a rationale of these considerations in the following paragraphs.

1. **Co-translational delivery.** During the translation process several proteins pass the membrane of the endoplasmic reticulum. We did not consider this phenomenon as protein translocation.

2. **Post-translational delivery.** During post-translational maturation processes several proteins change their localization in the cell, e.g. move from the endoplasmic reticulum to the Golgi apparatus where they may undergo additional post-translational modifications. We did not consider these protein movements as translocation as they are part of the post-translational maturation process and they are mostly not induced by a specific signal. We considered protein translocation as a phenomenon which happens after the protein reached its final destination and started to function.

3. **Protein translocation during the cell division process.** During mitosis or meiosis the subcellular membrane structure becomes reorganized. As one of the primary changes the integrity of the nuclear membrane is massively decreased, thus in these parts of the cell cycle nucleocytoplasmic translocation cannot be observed as in G0 phase. Therefore, protein translocations which were described during the cell division process were regarded with criticism and were added to the Translocatome database only if they had a functional meaning and not merely happened as a consequence of the cell division process.

4. **Protein movements related to downregulation.** Downregulation of several proteins (e.g. that of plasma membrane receptors) is achieved by their internalization to endocytotic vesicles and finally to the lysosome. We did not consider these phenomena as protein translocation.

5. **Protein movements related to passive diffusion.** If a protein is small enough it can passively transit through some subcellular membranes (where the best known example is the nuclear pore). This phenomenon results in a multicompartmental localization of these proteins. We, therefore, regarded the translocation of small-size proteins with special criticism.

## Supplementary Text S2. Additional considerations in the definition of protein translocation

1. **Definition of subcellular compartments.** We considered six subcellular compartments: cytoplasm, extracellular space, mitochondria, nucleus, membrane, secretory-pathway based on the compartmentalized protein-protein interaction database, ComPPI (http://comppi.linkgroup.hu, 1). We used these six major localizations because there is enough, comprehensive, high confidence information to create a localization-specific interactome (localization information is usually a result of a high-throughput method, therefore using more detailed localization items could increase the level of noise and create more bias in the data used). There are lower level functional organelles in cells (e.g.: proteasome, lysosome etc.), but their interactomes are not always well characterized and are usually not separated in system-level studies.

2. **Intra-compartmental translocations.** Most current methods thus do not make it possible to evaluate translocations happening inside major localizations. Thus, these "intra-compartmental" translocations where not included to the Translocatome database. However, we always saved the minor localization of translocating proteins, thus when systematic studies will provide a better resolution then the Translocatome database can be updated to cover these intra-compartmental translocations as well.

3. **Intra-compartmental moonlighting proteins.** Moonlighting proteins are known as proteins that are capable of executing different biochemical processes and are collected in the MoonProt 2.0 database (http://moonlightingproteins.org, 2). We considered moonlighting proteins as translocating proteins if their moonlighting involved more than a single subcellular organelle. There are proteins that have different function inside one subcellular organelle. As a specific example of the intra-compartmental translocation mentioned in point 2, these intra-compartmental moonlighting proteins were not included to our database.

4. **Extent of translocation.** We did not consider translocation as an 'all-or-none' phenomenon. Thus we included proteins as translocating proteins when the ratio of a protein in two given compartments changed substantially (e.g.: a protein that was mainly localized to the cytoplasm became mainly localized to the nucleus, however, this change was not complete, and there is still a small amount of the protein, which can be found in the cytoplasm).

5. **Participation in translocation.** Those papers, where "protein A" mediated the translocation of "protein B", or "protein A" interacted with the translocating "protein B" were considered only from the point of "protein B" and not regarding "protein A".

6. **Definition of regulated movement of a protein.** We collected translocating proteins participating in different cellular signalling mechanisms and responses, where the regulated movement of the protein was achieved by a signal. In addition, we considered regulated movement of the protein when the differing localization was achieved by the onset of a pathological condition (like e.g. cancer).

## Supplementary Text S3. Manual curation of translocating proteins

The manual curation team consisted of biochemists, bioinformaticians, cell biologists, molecular biologists and physicians. To ensure high quality manual curation every curated entry was assessed in the following way. After a paper was carefully reviewed by the primary curator it was subsequently reviewed by a senior member of the Translocatome team. If the data contained any inconsistencies or wasn't straightforwardly understandable in any way a discussion was initiated (involving at least one other member of the team) and the problematic information became corrected. The entry became accepted to the database only after every topic in question has been adequately cleared and the entry has been cross-checked by at least 3 independent experts in the process.

We give a flow chart of the manual curation process on Supplementary Figure S1. Our method of the curation process was similar to the methodology of previously published, manually curated databases like the MoonProt database (2).To find relevant scientific papers for building the gold standard positive dataset of 213 translocating proteins we used the PubMed and Google Scholar search engines. Key word-based searches resulted in a high number of hits as shown in Supplementary Table S1. After the initial reading process we realized that protein translocation has no unified definition and the word "*translocation*" is also used in a genomic sense to mark translocation of DNA segments or referring to the translocation of RNA on the ribosome. We have summarized our considerations for the definition of protein translocation in Supplementary Texts 1 and 2. This led to key word combinations of *protein translocation, nucleocytoplasmic translocation, nuclear translocation, cytoplasmic translocation and mitochondrial translocation*. We also realized that protein translocation is often referred as "*shuttling*" so we used this keyword, too. As shown on Table S1 these, more adequate key words and key word combinations significantly reduced the number of hits.

The manual curation process started on 25[th] October 2015 and finished at the end of 2017. During the browsing/filtering/selection process of the refined key word based searches we applied the following preferences and techniques:

- ➢ recent papers were preferentially checked, especially regarding review papers;
- ➢ several experimental papers were "traced back" using the relevant references of review papers and databases;
- ➢ papers appearing as "best match sorting" of PubMed searches and starting hits of Google Scholar searches were preferentially checked;
- ➢ key word searches were refined using word combinations in quotation marks;
- ➢ key word searches were refined using restrictions to "title" or "title/abstract";
- ➢ searches were considering only human proteins;
- ➢ only peer-reviewed papers were considered;
- ➢ papers on chromosomal, RNA or other types of translocations not related to protein translocation were not considered;
- ➢ the exclusion criteria summarized in Supplementary Text S1 were applied;
- ➢ the additional exclusion criteria summarized in Supplementary Text S2 (such as intra-compartmental translocation, intra-compartmental moonlighting, the necessity to have a regulated translocation movement, etc.) were applied;
- ➢ papers describing redundant information on previous hits were not considered.

As a result of this filtering/selection process we read over a hundred review papers and over a thousand pre-selected experimental papers submitting them to the manual curation process described in the starting paragraph of this Supplementary Text. Since we did not record the number of irrelevant articles from the start, we may only give estimates of the total papers read here. In addition, we also used direct, name-based searches for well-known translocating proteins, like nuclear hormone receptors, p53, EGFR, HIF1-alpha and other proteins, mentioned in review papers or in databases such as ComPPI (1) or MoonProt (2).

During the detailed reading process of the selected papers we first cross-checked all exclusion criteria mentioned in the screening process (including those listed in Supplementary Texts S1 and S2). We included only those proteins to our Core Data, where the papers were discussing not only the mere fact of their translocation but also the details of their translocation mechanism (e.g. its regulation or mechanism related to the structure of the translocating protein) based on experimental validation. However, these details of the translocation process were not discussed in equal depths in different articles. To give an information on the richness of data behind each of the 213 entries of our database we created the Data Complexity Score as described in the main text and in Supplementary Text S7.

We finished data collection when an increasing amount of former hits appeared in our new searches completing our Core Data having 213 human translocating proteins described in the 238 papers available at the link in the footnote[1]. This does not mean that the number of 213 manually curated translocating proteins of this Core Data (see here http://translocatome.linkgroup.hu/coredata) would give a complete list of human translocating proteins, as it is shown by Table 1 of the main text and Supplementary Table S4, where 11 out of the best 25 predictions turned to be already known translocating proteins not being part of our Core Data. This is one of the reasons why we created the Manual Curation Framework (MCF) to allow the continuous possibility to add further entries to the manually curated positive dataset. MCF entry suggestions will be reviewed with the same rigor described for the initial manual curation process. In addition, we complemented the manually curated information on translocating proteins with the efficient machine learning tool, XGBoost (3-5) to predict the translocation likelihood of approximately 13 000 human proteins as described in the main text and in Supplementary text S6 in detail.

---

[1] https://www.ncbi.nlm.nih.gov/pubmed?term=24491427+24434356+23435424+24074954+15071501+21900948+19210988+24113167+26449824+26079448+26520802+23955340+25989275+26406376+26056081+26602019+25684142+25823029+26648570+27013058+17308100+24667139+24695740+16772870+26942675+24725411+26516353+26387538+26549027+26470026+26499805+27292796+16390708+19524307+26827288+19808100+24735540+19718473+26969532+21419860+18760948+19854831+22583914+26181205+26379505+26643147+10749215+18974300+25589722+23814078+17123511+11818509+24954011+11679632+27641332+9765220+17786044+20498072+18075313+26559910+27572958+21469768+9822602+9039962+8643491+22345668+12628186+11517310+22534175+16640563+23707396+11231586+16162498+11756542+1985107+22684108+12649597+24841202+23333404+24013206+10477286+15809060+20132536+10512882+27099442+25609610+27346674+27913144+1532584+26891695+24573109+15485931+11301021+14505568+22639060+11043577+28149448+7925301+24324740+9672244+11705999+7782294+15084609+10187816+22186421+24628430+7748180+14733946+12432920+27496138+18636433+18716285+25835495+19815064+27875245+9636171+20826166+22482906+10397761+10811825+20674093+17516504+15496412+1312324+18204201+16782877+22932683+18515545+10973496+11559828+27428427+19216841+19160485+18045535+26319354+25862818+18337556+19651895+10744629+22832227+19503814+18238777+9679058+17289031+27245214+17157415+22178385+26462148+17535848+25635431+16110492+28235034+12115727+20346347+20308327+16807357+28515276+10547363+21157379+27721408+20511593+16403913+10508860+28552616+28504714+12011459+25893308+20651736+24196791+25961505+18334557+22885005+22824301+22391300+16215984+26738429+22692200+27251589+25701194+26500058+26358502+26633708+25882840+24965109+27272778+24784232+24489995+28811933+25893289+25164084+22473997+27462018+25755279+24959884+27941888+26900797+27510036+16375604+23128389+14715249+17595320+22623727+17967441+26651356+16792529+15314173+18973764+12015613+28812328+15728466+23979357+12191473+19114992+23384547+18061509+14769937+14652813+22296597+11925436+28863181+17218261+22989880+20148342+15951807+15803152+25926267+17003494+12809600+15623571+9660801+20605787+16818237+20937816+16951195+10811646+17560175+25299576+17645779%5Buid%5D&cmd=DetailsSearch&log$=activity

## Supplementary Text S4. Manual curation of non-translocating proteins

The negative dataset could not be searched using specific keywords as we did in the case of the gold standard positive translocating set, since there are no keywords to identify those proteins that are not translocating. Thus, we designed a different approach using the following considerations.

1. **Database filtering to reveal proteins having only a single minor localization.** Our first approach was to find proteins that only have only a single minor localization, which makes it rather unlikely that they translocate. In order to filter these proteins, we assessed the data of the compartmentalized protein-protein interaction (ComPPI, 1) the UniProt (6) and the Human Protein Atlas (HPA, 7) databases. We found 844 single minor localization proteins in ComPPI. From these proteins 98 had a single localization in UniProt. From these 98 entries 60 cases had the same localization in both databases. From these 60 proteins 18 had localization information in the HPA database. From these 18 proteins 11 proteins had the same localization in all 3 databases. These 11 proteins having the same localization in all 3 databases were included to the negative dataset. To extend this dataset we decided to include those proteins, which had the same, single minor localization in ComPPI (1) and HPA (7). To achieve this we started the filtration process with HPA. Here we found 1001 proteins which had only a single validated localization. We could map 996 SwissProt UniProt IDs to these proteins. We found 995 of these proteins in ComPPI, out of which 18 proteins were found, which had only a single minor localization in the complex dataset of ComPPI having close to 200 thousand localizations for human proteins. 7 of these proteins did not have localization data in UniProt (6). Thus, altogether this process resulted in 18 proteins that have only a single subcellular localization.

2. **Proteins with experimentally validated mono-compartment localization.** 29 proteins total, e.g. ASAT or GPT1. Note that among these proteins may have multi-compartmental localization achieved by high-throughput methods. Such proteins were not included to the search results described in point 1. Here we did include them, since we considered experimental validation more convincing than high-throughput results.

3. **Proteins with diffuse multicompartment localization.** Example of the 4 proteins with diffuse multicompartment localization: PEX12.

4. **Proteins anchored to the cytoskeleton.** Examples of the 16 proteins anchored to the cytoskeleton: titin or myosin.

5. **Docked proteins to the DNA or to the membrane.** Examples of the 26 DNA-bound proteins TAF1B or GATA3. Examples of the 64 membrane-bound proteins: Na/K ATPase or RYR.

Altogether these considerations resulted in 139 non-translocating proteins of the negative dataset (the 18 proteins of the search described in point 1 also appeared in later searches showing the integrity of the approach).

**Supplementary Text S5. Gene Ontology annotation of the proteins**

For each ComPPI protein (including the training set) we extracted the associated GO terms (8,9). However, in some cases a UniProt (6) entry is associated with a given GO term but not all of the ancestors of this GO term. As an example, the Merlin (P35240) protein entry is associated with the GO term "negative regulation of cell proliferation" (GO:0008285) but not its ancestor GO terms  "regulation of cell proliferation" (GO:0042127), "regulation of cellular process" (GO:0050794), "regulation of biological process" (GO:0050789), "biological regulation" (GO:0065007) and "biological process" (GO:0008150)." We solved this problem in the same way as described in earlier studies (5,10). We downloaded the basic version of the Gene Ontology database (with the database filename "go-basic.obo") and by walking upward in the GO hierarchy, we added all of the ancestor GO terms to the proteins. Note that "go-basic.obo" is guaranteed to have a hierarchical organization where annotations can be followed in the structure of the term-hierarchy. The final feature table contains 13 066 proteins and 21 020 binary (true/false) GO features.

## Supplementary Text S6. Prediction by the XGBoost machine learning method

**Labels of the binary classification.** The target variable (labels) of the binary classification has value 1 for the 160 manually curated translocating proteins ("translocating class") and value 0 for the 139 non-translocating proteins (non-translocating class).

**Feature selection on the training set.** The well-established and widely used XGBoost machine learning algorithm (3-5) is capable of selecting the most important features by building small decision trees of the most important features and gradually refining the models by adding new trees. We started the feature selection process using the set of 21 020 annotated GO features for the whole training set (n=299). We evaluated the XGBoost-selected feature sets by 5-fold cross-validation (in which we split the data into 5 random parts and in each round, used 4 parts to train and evaluated the prediction on the fifth part) and measured the area under the curve of the receiver operating characteristic curve (AUC,11). For every feature set, we repeated this process 100 times and computed the average AUC.

**Final prediction for all the 13 066 proteins.** This final feature set and the final model parameters was used to predict translocation for all the 13 066 proteins in the database. The selected features of the model with the best ROC AUC value is shown on Table 1 of the main text with their importance values calculated from the leaf-scores of the one-depth trees of the final XGBoost model (https://github.com/kerepesi/translocatome_ml/blob/master/Results/GO_features.csv-imp-n_est80-thr0.02-table.csv-add_degree_bridgeness.csv_Trees-n_est80-max_d1.txt). Using this feature set we calculated the Translocation Evidence Score characterizing the translocation probability of each of the 13 066 proteins in the database as described in a the main text.

**Evaluation measures for binary classification.** TP (true positive) is the number of positives that are predicted as positives. TN (true negative) is the number of negatives that are predicted as negatives. FP (false positive) is the number of negatives that are predicted as positives. FN (false negative) is the number of positives that are predicted as negatives. In our context "positive" means "translocating", "negative" means "non-translocation". Precision, recall (or true positive rate), fall-out (or false positive rate), F1 score and Matthews correlation coefficient were computed by Equations 1 to 5, respectively:

$$precision := \begin{cases} \dfrac{TP}{TP + FP}, & \text{if } TP + FP \neq 0, \\ 1, & \text{otherwise.} \end{cases} \quad (1)$$

$$recall := \dfrac{TP}{TP + FN}, \quad (2)$$

$$fall-out := \dfrac{FP}{TN + FP} \quad (3)$$

$$F1\ score := \begin{cases} \dfrac{2 \cdot precision \cdot recall}{precision + recall}, & \text{if } precision + recall \neq 0, \\ 0, & \text{otherwise.} \end{cases} \quad (4)$$

$$MCC := \begin{cases} \dfrac{TP \cdot TN - FP \cdot FN}{\sqrt{(TP+FP)(TP+FN)(TN+FP)(TN+FN)}}, & \text{if } (TP+FP)(TN+FN) \neq 0, \\ 0, & \text{otherwise.} \end{cases} \quad (5)$$

To evaluate the final prediction of the XGBoost method (3-5), we plotted the receiver operating characteristic curve (ROC, Supplementary Figure S4A). The performance of the model was defined as the area under the curve of the receiver operating characteristic curve (AUC, 11). The receiver operating characteristic curve (ROC) is defined by the point pairs of recall (or true positive rates) and fall-out (or false positive rates) at different threshold settings (11). We show the ROC curves, as well as precision-recall curves and Matthews correlation coefficient curves of 100 five-fold cross-validation runs on Figure 1C of the main text and Supplementary Figure S3, respectively, all using the final feature set (see Table 1 of the main text). All showed a high performance as discussed in the main text and legend related to Figure 1C and the legend of Supplementary Figure S3.

**Supplementary Text S7. Calculation of the Data Complexity Score (DCS)**

The data complexity score (DCS) characterizes the information content of the manually curated protein-associated data of the Translocatome database. DCS is calculated as a weighted measure. The following information can increase the DCS:

| Information category | Weight of the respective information |
|---|---|
| well described translocation mechanism | 3 |
| protein structural background of the translocation | 3 |
| known biological processes in the subcellular compartments | 2 |
| known interactions in the subcellular compartments | 1 |
| affected signalling pathway is discovered | 1 |
| known pathological function of the translocating protein | 1 |
| type of the pathological condition affected by translocation | 1 |
| exact pathology for which the translocation is an underlying factor is discovered | 1 |
| detection method of the cellular localizations is known | 1 |

After summing the weight values that are true for a given entry we divide this number with the maximal achievable DCS score of 23. Thus, DCS is a weighted and normalized average of the information categories, which can vary between the minimum of zero and the maximum of 1.

## Supplementary Text S8. Calculation of the F1 score

Contribution of the various GO-related and network-related features selected by the XGBoost machine learning algorithm (3-5) made us possible to define the Translocation Evidence Score (representing the likelihood of the translocation of a given protein) for all the 13 066 human proteins of Translocatome as described in the main text. The Translocation Evidence Score gave the possibility to define a cut-off value, below which proteins were considered as non-translocating. To define this cut-off value, we used the parameter, F1 score (also called as F-measure, see Equation 4 in Supplementary Text S6, and supplementary reference 12). To calculate the F1 score we defined true positive hits as the positive training set proteins that has been predicted as translocating, true negative hits as the negative training set proteins that has been predicted as negative, false positive hits as the negative training set proteins that has been predicted as translocating and false negative hits as the positive training set proteins that has been predicted as non-translocating proteins – as described in Supplementary Text S6 in more detail.

**Supplementary Text S9. Design and implementation of the Translocatome database.**

The Translocatome database has altogether 13 066 human protein entries. The core dataset is the 213 manually curated human translocating proteins. Our primary aim was to implement this database as a user-friendly application which is easily browsable, understandable and has powerful search and download options. The website is designed to satisfy the needs of different disciplines providing data addressing several aspects of the translocation phenomenon.

The database has an industry standard-level software design by a multidisciplinary team (database expert, informatician, bioinformatician, physician, graphic expert). The Translocatome website is designed as a client-server architecture: a NodeJS back-end serves the requests and fetches data from the MongoDB database, while the front-end, implemented with React.js displays this data in a user-friendly way.

We selected third-party tools and technologies that favour scientific reproducibility and open accessibility, including the Ubuntu Linux 16.04 operating system (http://ubuntu.com), the nginx HTTP server (http://nginx.org), the MongoDB 3.2 database server (https://www.mongodb.com), the git version control system (http://git-scm.com), the NodeJS 6.10 (https://nodejs.org/en) Javascript server runtime with the Express.js 4.15 framework (http://expressjs.com) and the React.js 15.4 (https://reactjs.org) JavaScript framework.

The dataset can be browsed and searched by various user preferences. Protein names are auto-completed using UniProt (6) accession numbers. The advanced search option gives the choice of translocation direction (from one selected cellular compartment to another), as well as the range of Translocation Evidence Score and Data Complexity Score. During searching a NodeJS script automatically generates a downloadable version of the current data. Furthermore, we provide pre-defined download sets.

The files forming the base of the Translocatome database can be generated with a simple SQL query from our previously developed, MySQL-based compartmentalized protein-protein interaction database (ComPPI, 1). This feature connects the two databases. Thus the improvement of the subcellular localization and interactome data can be easily translated to regular updates of the Translocatome database giving improved protein translocation probability values.

# Supplementary Tables

### Supplementary Table S1. Number of hits in PubMed and Google Scholar searches

| Key words | Unrestricted search PubMed/Google Scholar* | Reviews only available in PubMed | Filtered to human available in PubMed |
|---|---|---|---|
| translocation | 297 319/133 000 *(329 989/154 000)\*\** | 32 602 *(36 096)* | 144 728 *(160 749)* |
| protein translocation | 251 037/99 000 *(278 900/109 000)\*\** | 26 922 *(29 802)* | 116 726 *(130 928)* |
| shuttling | 3420/21 000 *(4001/23 000)\*\** | 402 *(466)* | 1743 *(1972)* |
| nucleocytoplasmic translocation | 2711/4000 *(3016/8000)\*\** | 347 *(403)* | 1551 *(1740)* |
| nuclear translocation | 43 891/64 000 *(50 023/72 000)\*\** | 3568 *(3955)* | 25 130 *(28506)* |
| cytoplasmic translocation | 84 923/35 000 *(93 122/45 000)\*\** | 8902 *(9799)* | 38 047 *(42 454)* |
| mitochondrial translocation | 25 239/24 000 *(28 653/48 000)\*\** | 2908 *(3318)* | 10 409 *(12 135)* |
| **Total** | **299 216/133 000** ***(332 304/154 000)\*\**** | **32 858** *(36 398)* | **145 528** *(161 674)* |

To find relevant scientific papers for building the gold standard positive dataset of 213 translocating proteins we used the PubMed and Google Scholar search engines searching for the keywords: *translocation, protein translocation, shuttling, nucleocytoplasmic translocation, nuclear translocation, cytoplasmic translocation and mitochondrial translocation.* Articles found were manually curated using the definition and the exclusion criteria as discussed in Supplementary Text S3. This table shows the number of PubMed and Google Scholar search results for each keyword separated by a slash. In the first column the results of the unrestricted search are shown. We also included the number of search results if we filtered for only review articles or only human studies. The manual curation process started on the 25.09.2015. and the numbers show the results restricted until that time. Italic numbers in parentheses refer to the number of papers restricted until the end of the manual curation process which was on 31 December 2017. For the assembly of the data in this table both databases were accessed on 4[th] October 2018.

*Google Scholar hits were rounded to thousands. The total of Google Scholar counts refers to the search of "translocation" since the addition of "OR shuttling" term did not improve the number of hits. Without any restrictions Google Scholar gives over 2 million hits for "translocation".

**Italic numbers in parentheses refer to the number of papers restricted until the end of the manual curation process which was on 31 December 2017.

## Supplementary Table S2. Positive training set

| UniProt AC | Gene name | UniProt AC | Gene name | UniProt AC | Gene name | UniProt AC | Gene name |
|---|---|---|---|---|---|---|---|
| Q8IZP0 | ABI1 | Q9UER7 | DAXX | P17936 | IGFBP3 | P12272 | PTHLH |
| P00519 | ABL1 | Q08211 | DHX9 | P24593 | IGFBP5 | P49023 | PXN |
| P42684 | ABL2 | Q13316 | DMP1 | P24592 | IGFBP6 | P20339 | RAB5A |
| Q9NR19 | ACSS2 | P26358 | DNMT1 | Q92985 | IRF7 | P63000 | RAC1 |
| P35869 | AHR | Q01094 | E2F1 | Q92830 | KAT2A | Q06609 | RAD51 |
| Q9GZX7 | AICDA | Q15029 | EFTUD2 | Q92831 | KAT2B | P10276 | RARA |
| P55008 | AIF1 | P00533 | EGFR | P01116 | KRAS | Q08999 | RBL2 |
| O95831 | AIFM1 | P18146 | EGR1 | Q9UN81 | L1RE1 | P18754 | RCC1 |
| P31749 | AKT1 | Q9Y6B2 | EID1 | Q14847 | LASP1 | O94761 | RECQL4 |
| P09917 | ALOX5 | P06730 | EIF4E | P02545 | LMNA | Q04206 | RELA |
| Q8NAG6 | ANKLE1 | Q15717 | ELAVL1 | Q02750 | MAP2K1 | P51449 | RORC |
| Q7Z6G8 | ANKS1B | P78545 | ELF3 | P28482 | MAPK1 | P62829 | RPL23 |
| P10275 | AR | P06733 | ENO1 | P27361 | MAPK3 | P23396 | RPS3 |
| P10398 | ARAF | Q15303 | ERBB4 | P29966 | MARCKS | Q13950 | RUNX2 |
| P15336 | ATF2 | P03372 | ESR1 | O00255 | MEN1 | Q01105 | SET |
| P18848 | ATF4 | P09038 | FGF2 | P15941 | MUC1 | O00141 | SGK1 |
| P35670 | ATP7B | Q12778 | FOXO1 | P19878 | NCF2 | Q96EB6 | SIRT1 |
| P54252 | ATXN3 | O43524 | FOXO3 | P19338 | NCL | P14672 | SLC2A4 |
| P15291 | B4GALT1 | P04406 | GAPDH | O95644 | NFATC1 | Q15797 | SMAD1 |
| Q16611 | BAK1 | P10071 | GLI3 | Q14934 | NFATC4 | P48431 | SOX2 |
| Q07812 | BAX | O75496 | GMNN | Q16236 | NFE2L2 | P48436 | SOX9 |
| Q07817 | BCL2L1 | P04899 | GNAI2 | P19838 | NFKB1 | Q9NYA1 | SPHK1 |
| O15392 | BIRC5 | P56524 | HDAC4 | Q00653 | NFKB2 | Q12772 | SREBF2 |
| P38398 | BRCA1 | Q9UQL6 | HDAC5 | P29474 | NOS3 | Q07955 | SRSF1 |
| P27797 | CALR | Q16665 | HIF1A | P04150 | NR3C1 | P42224 | STAT1 |
| O14936 | CASK | P16403 | HIST1H1C | Q6X4W1 | NSMF | P40763 | STAT3 |
| P14635 | CCNB1 | P52789 | HK2 | Q8TAK6 | OLIG1 | Q13043 | STK4 |
| Q16589 | CCNG2 | P09429 | HMGB1 | P49585 | PCYT1A | P32856 | STX2 |
| P50750 | CDK9 | P09651 | HNRNPA1 | O15534 | PER1 | Q15561 | TEAD4 |
| P46527 | CDKN1B | O14979 | HNRNPDL | P00558 | PGK1 | O14746 | TERT |
| P49918 | CDKN1C | P52597 | HNRNPF | P14618 | PKM | P36897 | TGFBR1 |
| P17676 | CEBPB | P31943 | HNRNPH1 | P00749 | PLAU | P21980 | TGM2 |
| O14757 | CHEK1 | P61978 | HNRNPK | P37231 | PPARG | P10828 | THRB |
| Q99828 | CIB1 | Q00839 | HNRNPU | Q00005 | PPP2R2B | Q9BSI4 | TINF2 |
| P49759 | CLK1 | P01112 | HRAS | Q15172 | PPP2R5A | P62328 | TMSB4X |
| Q16526 | CRY1 | P07900 | HSP90AA1 | Q14738 | PPP2R5D | P04637 | TP53 |
| P56545 | CTBP2 | P08238 | HSP90AB1 | P30041 | PRDX6 | Q13114 | TRAF3 |
| P35222 | CTNNB1 | P11142 | HSPA8 | P17612 | PRKACA | Q9NSU2 | TREX1 |
| Q14247 | CTTN | Q02363 | ID2 | P78527 | PRKDC | P68543 | UBXN2A |
| P99999 | CYCS | P18065 | IGFBP2 | P60484 | PTEN | P46937 | YAP1 |

Supplementary Table S2 contains the 160 manually curated translocating proteins that translocate under physiological conditions and thus were selected as elements of the positive training set.

## Supplementary Table S3. Negative training set

| UniProt AC | Gene name | UniProt AC | Gene name | UniProt AC | Gene name | UniProt AC | Gene name |
|---|---|---|---|---|---|---|---|
| A6NGB9 | WIPF3 | P08922 | ROS1 | P48764 | SLC9A3 | Q92824 | PCSK5 |
| A8MQ14 | ZNF850 | P0C024 | NUDT7 | P49790 | NUP153 | Q92908 | GATA6 |
| A8MUV8 | ZNF727 | P0CJ78 | ZNF865 | P51946 | CCNH | Q96HA9 | PEX11G |
| B4DU55 | ZNF879 | P10635 | CYP2D6 | P54277 | PMS1 | Q96LT4 | SAMD8 |
| C9JN71 | ZNF878 | P10721 | KIT | P60709 | ACTB | Q96MW1 | CCDC43 |
| E7ETH6 | ZNF587B | P11055 | MYH3 | P62068 | USP46 | Q99595 | TIMM17A |
| O00391 | QSOX1 | P12883 | MYH7 | P68133 | ACTA1 | Q99705 | MCHR1 |
| O00623 | PEX12 | P13533 | MYH6 | Q12824 | SMARCB1 | Q99965 | ADAM2 |
| O14576 | DYNC1I1 | P13535 | MYH8 | Q12931 | TRAP1 | Q9BRQ3 | NUDT22 |
| O14925 | TIMM23 | P13569 | CFTR | Q13423 | NNT | Q9BRT6 | LLPH |
| O14964 | HGS | P14416 | DRD2 | Q14108 | SCARB2 | Q9BWM7 | SFXN3 |
| O14975 | SLC27A2 | P16278 | GLB1 | Q14249 | ENDOG | Q9HCE1 | MOV10 |
| O14983 | ATP2A1 | P16473 | TSHR | Q15388 | TOMM20 | Q9NRP2 | CMC2 |
| O60341 | KDM1A | P16581 | SELE | Q15722 | LTB4R | Q9NS69 | TOMM22 |
| O60563 | CCNT1 | P17174 | GOT1 | Q16678 | CYP1B1 | Q9NUJ7 | PLCXD1 |
| O75027 | ABCB7 | P18847 | ATF3 | Q16822 | PCK2 | Q9NZ42 | PSENEN |
| O75192 | PEX11A | P18859 | ATP5J | Q16878 | CDO1 | Q9NZ52 | GGA3 |
| O75694 | NUP155 | P19367 | HK1 | Q53T94 | TAF1B | Q9NYW8 | RBAK |
| O94822 | LTN1 | P19404 | NDUFV2 | Q71U36 | TUBA1A | Q9P0Z9 | PIPOX |
| O94826 | TOMM70 | P20309 | CHRM3 | Q7Z406 | MYH14 | Q9UH99 | SUN2 |
| O94901 | SUN1 | P20585 | MSH3 | Q7Z412 | PEX26 | Q9UHC1 | MLH3 |
| O95182 | NDUFA7 | P21439 | ABCB4 | Q86Y39 | NDUFA11 | Q9UHD2 | TBK1 |
| O95198 | KLHL2 | P21817 | RYR1 | Q86YV0 | RASAL3 | Q9UJ83 | HACL1 |
| O96008 | TOMM40 | P22303 | ACHE | Q8IWY9 | CDAN1 | Q9UJM8 | HAO1 |
| P01106 | MYC | P23771 | GATA3 | Q8IXM3 | MRPL41 | Q9UKU7 | ACAD8 |
| P03923 | MT-ND6 | P24298 | GPT | Q8N9L9 | ACOT4 | Q9UKX2 | MYH2 |
| P04629 | NTRK1 | P24752 | ACAT1 | Q8N9W6 | BOLL | Q9UL17 | TBX21 |
| P05023 | ATP1A1 | P35548 | MSX2 | Q8NGJ1 | OR4D6 | Q9UQ90 | SPG7 |
| P05026 | ATP1B1 | P35558 | PCK1 | Q8TD30 | GPT2 | Q9Y259 | CHKB |
| P05412 | JUN | P35579 | MYH9 | Q8TES7 | FBF1 | Q9Y2Q9 | MRPS28 |
| P05496 | ATP5G1 | P35580 | MYH10 | Q8WVN6 | SECTM1 | Q9Y2W1 | THRAP3 |
| P06276 | BCHE | P35749 | MYH11 | Q8WVX9 | FAR1 | Q9Y3A0 | COQ4 |
| P06400 | RB1 | P36542 | ATP5C1 | Q8WXD0 | RXFP2 | Q9Y584 | TIMM22 |
| P07949 | RET | P40939 | HADHA | Q8WZ42 | TTN | Q9Y5J7 | TIMM9 |
| P08684 | CYP3A4 | P48058 | GRIA4 | Q92621 | NUP205 | | |

Supplementary Table S3 contains the 139 manually collected non-translocating proteins that were collected based on the considerations discussed in Supplementary Text S4 and form the negative training set.

## Supplementary Table S4. Occurrence of the 11 top high-confidence translocating proteins not part of our Core Data in PubMed searches

| Protein name | UniProt ID | PubMed ID | PubMed search term | | | | | | |
|---|---|---|---|---|---|---|---|---|---|
| | | | "translocation" | "protein translocation" | "shuttling" | "nucleo-cytoplasmic translocation" | "nuclear translocation" | "cytoplasmic translocation" | "mitochondrial translocation" |
| NF2 | P35240 | 24726726 20178741 | **21** 29709009 | 0 | 1 | 0 | **4** 29709009 | 1 | 0 |
| TULP3 | O75386 | 11375483 | **3** 11375483 | 0 | 0 | 0 | 0 | 0 | 0 |
| SNCA | P37840 | 18440504* | **5** | 0 | 2 | 0 | 0 | 0 | 0 |
| FGFR2 | P21802 | 15654655 | **32** 16365892 15654655 | 0 | 0 | 0 | **5** 16365892 15654655 | 0 | 0 |
| MTOR | P42345 | 11114166* | **14** 29512299 28694500 21822208 | 0 | **2** 26097872 | 0 | **3** 21822208 | 0 | 0 |
| GSK3B | P49841 | 17438332* | **7** | 0 | 0 | 0 | 0 | 0 | 0 |
| EIF6 | P56537 | 21084295 | **2** 21084295 | 0 | **1** 21084295 | 0 | 0 | 0 | 0 |
| HDAC1 | Q13547 | 15897453 20037577 27669993 | **16** 24658119 | 1 | 2 | 0 | **9** 24658119 | 0 | 0 |
| CARM1 | Q86X55 | 19843527 17848568 19208762 | **4** 29681515 19843527 | 0 | 1 | 0 | **2** 29681515 | 0 | 0 |
| CUL1 | Q13616 | 26068074** 11027288 | **13** 21247897 | 0 | 0 | 0 | **4** 21247897 | 1 | 0 |
| RARB | P10826 | 19501957* | **5** | 0 | 0 | 0 | 0 | 0 | 0 |

Supplementary Table S4 contains the number of papers **in boldface** retrieved from Title/Abstract (in case of MTOR and HDAC1 Title/Title and HDAC1/Title, respectively) restricted PubMed searches. Nine digit PubMed IDs refer to those papers among these, where experimental evidence of translocation was listed. Experimental evidence for the translocation of three proteins (SNCA, GSK3B and RARB) were not found in keyword searches of their UniProt names, since these proteins were mentioned by different name versions in the papers describing the experimental evidence for their translocation.
*The name of the protein was not the same as its UniProt name in the paper.
**Note that this paper refers to an "SCF Ubiquitin Ligase Complex" which includes CUL1.

## Supplementary Figures

### Supplementary Figure S1. Workflow of the data acquisition process

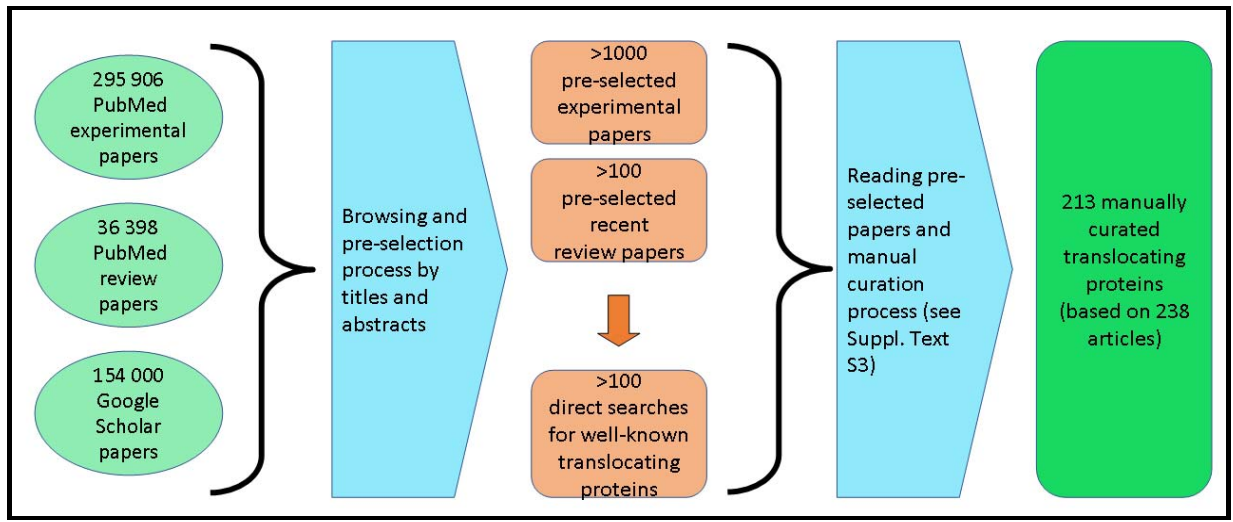

The figure shows the flow-chart of our manual curation process, which started on 25$^{th}$ October 2015 and finished at the end of 2017. The curation process was similar to the methodology of previously published, manually curated databases like the MoonProt database (2). We started from a pool of 295, 36 and 154 thousand PubMed experimental papers, PubMed reviews and Google Scholar papers, respectively, as shown on Supplementary Table S1 in detail. We especially considered recent review papers as references to more detailed searches and studies. The browsing/pre-selection process had the preferences described in Supplementary Text S3 in detail. As a result of this process we read more than a thousand research papers and over hundred reviews. We also made over a hundred direct searches for translocating proteins suggested by reviews and databases like ComPPI (1) or MoonProt (2). Applying the selection criteria of Supplementary Texts 1 through 3 during the careful manual curation process described in the starting paragraph of Supplementary Text S3, we finally selected the 238 papers which describe experimental evidence of the translocation of 213 human proteins forming the Core Data of our database (see here: http://translocatome.linkgroup.hu/coredata).

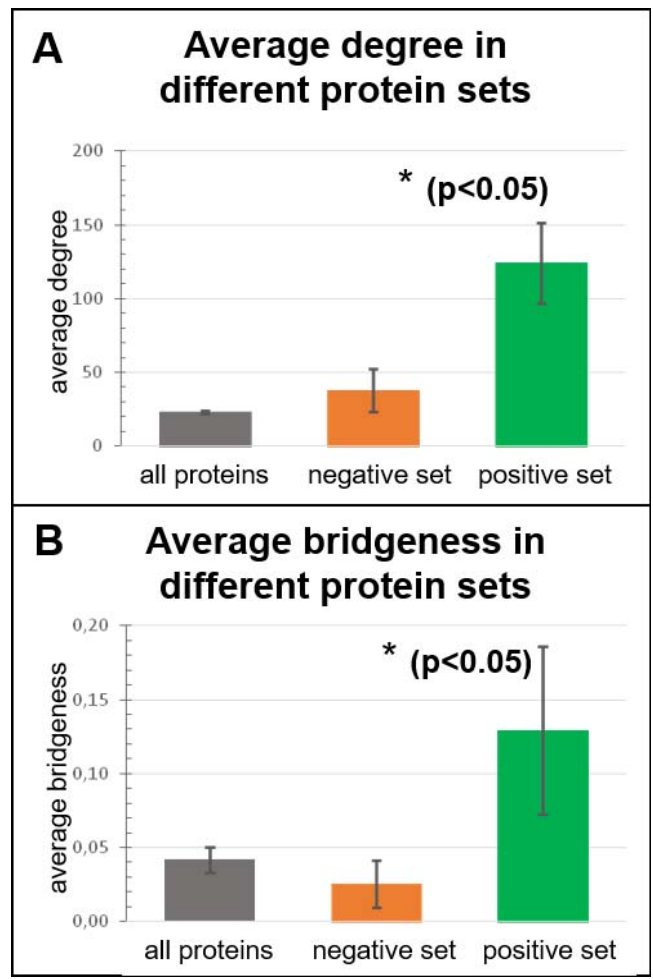

**Supplementary Figure S2. Degree and bridgeness of all proteins, as well as the positive and negative training sets** The figure shows the mean ± SD of the degree (Panel **A**) and the bridgeness (Panel **B**) values of all the 13 066 proteins as well as the 160 and 139 proteins of the positive and negative training sets, respectively. Using the giant component of the compartmentalized protein-protein interaction network (ComPPI)-derived human interactome having 151 889 interactions (all downloadable from here: http://translocatome.linkgroup.hu/download, 1) the degree and bridgeness were calculated by CytoScape (13) and by its network module determination plug-in, ModuLand (14), respectively. Panel A shows that the average degree is 23.2, 37.9 and 124.1 of all proteins, the negative and the positive training sets, respectively. The average degree of the positive set is significantly higher than the degree of the other two sets ($p<0.05$, Student's two tailed t-test). Panel B shows that the average bridgeness is 0.04, 0.03 and 0.13 of all proteins, the negative and positive training sets, respectively. The average bridgeness of the positive set is significantly higher than the bridgeness of the other two sets ($p<0.05$, Student's two-tailed t-test).

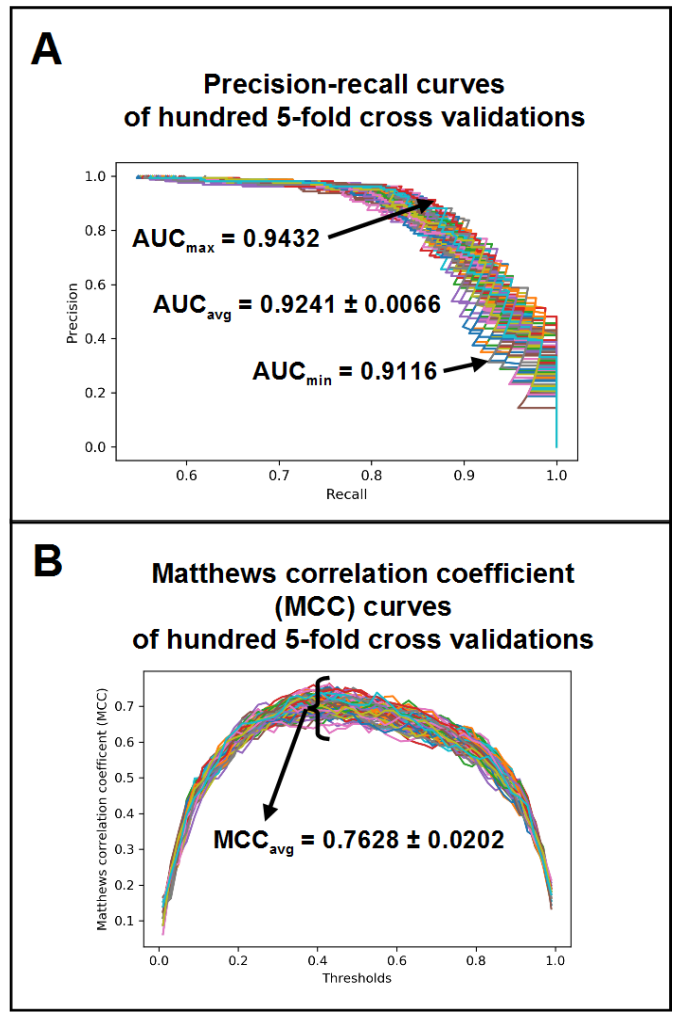

**Supplementary Figure S3. Performance of the widely-used XGBoost machine learning method on the final feature set evaluated by precision-recall and Matthews correlation coefficient curves of hundred 5-fold cross validations**

Each of the 100 different precision-recall (defined as Equations 1 and 2 of Supplementary Text S6) curves (panel A) and each of the 100 different Matthews correlation coefficient (MCC; defined as Equation 5 of Supplementary Text S6) curves (panel B) belong to a different 5 fold cross-validation run on the training set (containing 160 translocating and 139 non-translocating proteins). MCC values of the hundred 5-fold cross-validations were plotted as the function of threshold values dissecting the total range of MCC values to 50 equal segments. 5-fold cross validation runs were identical to those, whose receiver-operating characteristic (ROC) curves were shown on Figure 1C of the main text. ROC, precision-recall and MCC values are the generally suggested evaluation measures of machine learning methods (see e.g. in Refs. 12 and 15). In these 5-fold cross validation runs the well-established XGBoost machine learning method (3-5) used the final feature set (as shown on Table 1 of the main text) selected as described in the main text and Supplementary Text S6. The minimum, maximum and average of the area under the precision-recall curve values were 0.9116, 0.9432 and 0.9291 (±0.0066 standard deviation), respectively. The minimum, maximum and average of the maximum MCC values were 0.6627, 0.7628 and 0.7202 (±0.0202 standard deviation), respectively. MCC values range between -1.0 and +1.0, where all MCC values higher than zero mean better predictions than random choice, and MCC=1 means a perfect prediction (12,15,16).
20

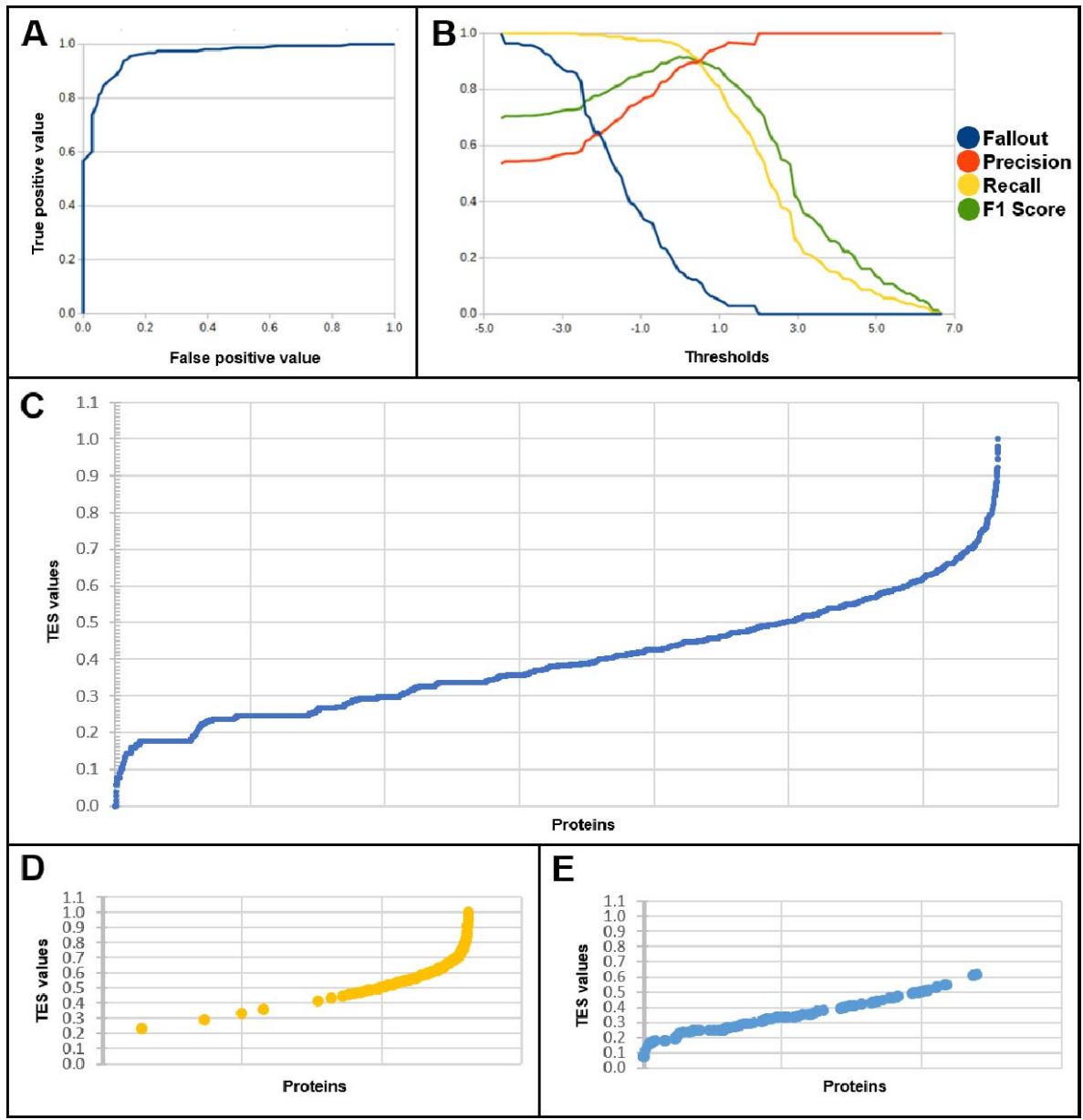

**Supplementary Figure S4. Statistical parameters of the final prediction and translocation evidence-(TES)-scores** Panel **A** shows the ROC curve for the prediction of the final model for the whole training set using the well-established, widely used XGBoost machine learning method (3-5, for more details see Supplementary Text S6). Panel **B** shows how the statistical parameters fallout, precision, recall and F1 score (see Equations 1 to 4 of Supplementary Text S6 and Supplementary Text S8) change their values as a function of the non-normalized Translocation Evidence Score (TES) values (for its detailed explanation see the main text). Panels **C, D** and **E** show the distribution of the TES values of all proteins, as well as those of the positive and negative training sets, respectively. Note that most of the positive training set entries have high TES values. On the contrary, most of the negative training set proteins have low TES values.

[Table image showing spreadsheet columns A through X with protein data rows for Q16666, Q9GZX7, P18065, Q16611, Q07812, O15392, P17612, P04637]

**Supplementary Figure S5. Data structure of manually curated translocating protein download files** Pre-defined download option files are in comma separated format (.csv), which can be opened using a spreadsheet. The figure shows a representative section from the "Manually curated translocating proteins" file. Each row represents a protein, with its data next to it. Each column contains the type of information marked in the upper top row. These explanations of the abbreviations of data types are:

- **UniProt AC**: To identify every protein we use their UniProt accession number (6). You can find this number on the UniProt website (https://www.uniprot.org).
- **Gene name:** Here we use the UniProt (6) Gene name. As a default we use the primary gene name. Some of the entries has more than one gene names separated with a pipe symbol.
- **Protein name:** This is the full name of a protein from the UniProt database (6).
- **Reference:** List of the PubMed IDs of the articles that discusses the information added to the latter cells. If more than one, they are separated with a pipe symbol.
- **Localization A:** For every localization data we use a specific type of input (called subdocument). See the example of the IFI16 protein:

    *{"major loc": ["nucleus (GO:0005634)"],"*
    *minor loc":"nucleoplasm (GO:0005654)",*
    *"comments": ["write your comments here"]}*

    Every localization item is stored according to the Gene Ontology naming convention (8,9). We use six major localizations: Cytoplasm, Nucleus, Mitochondrion, Extracellular space, Secretory-pathway, Membrane. "Minor loc" refers to a subcompartment of the six major localizations, if mentioned in the reference article. If there is any, comments are added to the "comments" section.

- **Localization B:** see at "Localization A"
- **Localization C:** see at "Localization A"
- **Translocation mechanism A-B:** Briefly summarizes data how the translocation occurs between localizations A and B.
- **Translocation mechanism B-C:** Briefly summarizes data how the translocation occurs between localizations B and C.
- **Structural information:** Briefly summarizes the structural properties of a given protein that are relevant for its translocation.
- **Biological process A:** List of the biological processes in localization A related to the protein stored according to the Gene Ontology naming convention (8,9), separated with the pipe symbol.
- **Biological process B:** see "Biological process A"
- **Biological process C:** see "Biological process A"
- **Interactions A:** The interacting protein names of corresponding entry in Localization A, specifically mentioned in the scientific paper cited in "Reference". If more than one, separated with the pipe symbol.
- **Interactions B**: The same as "Interactions A", but in Localization B.
- **Interactions C:** The same as "Interactions A", but in Localization C.
- **Signalling pathway**: The KEGG-based (17) naming convention of the signalling pathway associated with the protein. The following roles of the protein (often after posttranslational modification) are possible: inhibiting/blocking, enhancing, modifying, etc.

- **Pathological role:** If the cited scientific paper in "Reference" mentions anything about a pathological situation the translocation playing role in or leading to, it is to be marked here in as a pathophysiological factor (such as: tumorigenesis, angiogenesis, etc.).
- **Disease group**: Broad and general group of pathological states where the translocation was observed or occurs frequently (such as cancer).
- **Exact disease**: Exact disease, where the translocation was observed or occurs frequently.
- **Localization detection method**: The name of the biochemical method for observing the translocation (usually in a broader, general manner).

If any item of data is not available for the protein, the "N/A" marker is used.

|   | A | B | C | D | E | F | G | H | I | J |
|---|---|---|---|---|---|---|---|---|---|---|
| 1 | uniprot ac | gene nam | protein na | degree | betweenr | bridgenes | localizatic | translocation evidence score | | |
| 2 | A1E959 | ODAM | Odontoge | 3.0 | 1.01e-06 | 2.0e-06 | cytosol:0. | 0.6305 | | |
| 3 | A7KAX9 | ARHGAP3 | Rho GTPas | 24.0 | 2.606e-05 | 0.000446 | cytosol:0. | 0.7187 | | |
| 4 | O00148 | DDX39A | ATP-depe | 36.0 | 0.0001111 | 0.611863 | cytosol:0. | 0.633 | | |
| 5 | O00170 | AIP | AH recept | 28.0 | 0.0002386 | 0.015102 | cytosol:0. | 0.6614 | | |
| 6 | O00213 | APBB1 | Amyloid k | 71.0 | 0.0003435 | 0.328265 | cytosol:0. | 0.702 | | |
| 7 | O00221 | NFKBIE | NF-kappa- | 50.0 | 5.157e-05 | 0.001217 | cytosol:0. | 0.6614 | | |
| 8 | O00231 | PSMD11 | 26S prote | 107.0 | 0.0003185 | 0.006785 | cytosol:0. | 0.6338 | | |
| 9 | O00232 | PSMD12 | 26S prote | 79.0 | 6.76e-05 | 0.009573 | cytosol:0. | 0.7454 | | |
| 10 | O00233 | PSMD9 | 26S prote | 20.0 | 3.417e-05 | 0.000148 | cytosol:0. | 0.6298 | | |

**Supplementary Figure S6. Data structure of pre-defined download option files**

Pre-defined download options are in comma separated format (.csv), which can be opened using a spreadsheet. The figure shows the representative view of the following files: "High-confidence translocating proteins", "Low-confidence translocating proteins", "Non-translocating proteins predicted by machine learning", "Whole dataset of the Translocatome". Note that "Manually curated non-translocating proteins" file differs from the above mentioned ones, because columns G and H are excluded. Please note that the high-confidence, low-confidence and non-translocating datasets do not contain the manually curated proteins, since the extensive data structure of the latter is available in different download formats (see Supplementary Figure S5). Each row represents a protein, with its data next to it. Each column contains the type of information marked in the upper top row. These explanations of the abbreviations of data types are:

- **UniProt AC:** To identify every protein we use their UniProt accession number (6). You can find this number on the UniProt website (https://www.uniprot.org).
- **Gene name:** Here we use the UniProt (6) Gene name. As a default we use the primary gene name. Some of the entries has more than one gene names separated with a pipe symbol.
- **Protein name:** This is the full name of a protein from the UniProt database (6).
- **Degree**: Number of interacting partners of the given protein, based on the human protein-protein interaction network constructed from the data of the ComPPI database (1). The measure was calculated using the CytoScape program (13).
- **Betweenness centrality:** the score represents the number of the shortest paths that pass through the protein. Higher betweenness centrality represents a higher ability to control the network, since more information passes through the protein. This centrality measure was not used in the machine learning process, since it did not characterize well the positive and negative training sets (data not shown). The measure was calculated using the CytoScape program (13).
- **Bridgeness:** Bridges are nodes which connect different network modules, i.e. large protein complexes in protein-protein interaction networks (18). Since translocating proteins often have different associating partners in their different locations they are often forming bridges in the protein-protein interaction network. The measure was calculated using the ModuLand CytoScape plug-in (14).
- **Localizations with localization score from ComPPI database** (excuded from "Manually curated non-translocating proteins"): The score gives the probability of a given protein to be found in a certain compartment, the number is imported from the ComPPI database (1).
- **Translocation evidence score** (excluded from "Manually curated non-translocating proteins"): The TES scores gives the likelihood of the protein to translocate as defined and discussed in detail in the main text.

| | A | B | C | D | E | F | G | H | I | J | K | L | M | N | O |
|---|---|---|---|---|---|---|---|---|---|---|---|---|---|---|---|
| 1 | Protein_A | Naming_C | Synonyms | Taxonomy | Protein_B | Naming_C | Synonyms | Taxonomy | Interactio | Interactio | Interactio | Interactio | Interactio | Data_Source | |
| 2 | O75173 | UniProtKE | O75173\|A | 9606 | P01011 | UniProtKE | P01011\|SE | 9606 | 0.9968744 | two-hybri | HPRD | N/A | 16099106 | ComPPI_v1.1 | |
| 3 | O75173 | UniProtKE | O75173\|A | 9606 | Q96GW7 | UniProtKE | Q96GW7\| | 9606 | 0.9997399 | in vitro(Ex | HPRD | N/A | 10986281 | ComPPI_v1.1 | |
| 4 | O75173 | UniProtKE | O75173\|A | 9606 | P01009 | UniProtKE | P01009\|SE | 9606 | 0.9991217 | two-hybri | HPRD\|HPI | N/A | 16099106 | ComPPI_v1.1 | |
| 5 | O75173 | UniProtKE | O75173\|A | 9606 | P00738 | UniProtKE | P00738\|H | 9606 | 0.9994462 | two-hybri | HPRD | N/A | 16099106 | ComPPI_v1.1 | |
| 6 | O75173 | UniProtKE | O75173\|A | 9606 | P16112 | UniProtKE | P16112\|A | 9606 | 0.9987715 | protease a | MatrixDB | N/A | 19744558 | ComPPI_v1.1 | |
| 7 | O75173 | UniProtKE | O75173\|A | 9606 | P35625 | UniProtKE | P35625\|TI | 9606 | 0.9996408 | protease a | MatrixDB | N/A | 19643179 | ComPPI_v1.1 | |
| 8 | O75173 | UniProtKE | O75173\|A | 9606 | Q05516 | UniProtKE | Q05516\|ZI | 9606 | 0.24 | physical ir | BioGRID\|I | N/A | 21988832 | ComPPI_v1.1 | |
| 9 | O75173 | UniProtKE | O75173\|A | 9606 | Q96MU7 | UniProtKE | Q96MU7\| | 9606 | 0.0 | physical ir | BioGRID\|I | N/A | 21988832 | ComPPI_v1.1 | |
| 10 | O75173 | UniProtKE | O75173\|A | 9606 | Q15717 | UniProtKE | Q15717\|EI | 9606 | 0.0 | physical ir | BioGRID | N/A | 19322201 | ComPPI_v1.1 | |
| 11 | O75173 | UniProtKE | O75173\|A | 9606 | P09958 | UniProtKE | P09958\|FU | 9606 | 0.9973007 | in vivo(Ex | HPRD | N/A | 14744861 | ComPPI_v1.1 | |
| 12 | O75173 | UniProtKE | O75173\|A | 9606 | O60687 | UniProtKE | O60687\|SI | 9606 | 0.9994452 | two-hybri | HPRD\|HPI | N/A | 18718938 | ComPPI_v1.1 | |
| 13 | P16112 | UniProtKE | P16112\|A | 9606 | P22894 | UniProtKE | P22894\|M | 9606 | 0.9979114 | in vitro(Ex | HPRD | N/A | 8216228 | ComPPI_v1.1 | |
| 14 | P16112 | UniProtKE | P16112\|A | 9606 | P50281 | UniProtKE | P50281\|M | 9606 | 0.9992210 | in vivo(Ex | HPRD\|HPI | N/A | 11854269 | ComPPI_v1.1 | |
| 15 | P16112 | UniProtKE | P16112\|A | 9606 | P51884 | UniProtKE | P51884\|LL | 9606 | 0.9993775 | physical ir | BioGRID\|I | N/A | 15505028 | ComPPI_v1.1 | |
| 16 | P16112 | UniProtKE | P16112\|A | 9606 | Q9UHI8 | UniProtKE | Q9UHI8\|A | 9606 | 0.9964003 | in vitro(Ex | HPRD | N/A | 12054629 | ComPPI_v1.1 | |

**Supplementary Figure S7. Data structure of the protein-protein interaction network downloadable file** Pre-defined download options are in comma separated format (.csv), which can be opened using a spreadsheet. The figure shows the representative view of the following file: "Protein-protein interaction network of the Translocatome". The following data columns (the entries are separated by |) can be seen:

- **Protein A:** UniProt (6) accession of interactor protein A.
- **Naming Convention A:** naming convention for the interactor protein A. Only Swiss-Prot entries were used, so the value is "UniProtKB/Swiss-Prot/P" for every entry.
- **Synonyms A:** list of the protein name synonyms for the interactor protein A
- **Taxonomy ID A:** the taxonomy ID of the interactor protein A, since all the protein are human, the ID is 9606 for every entry.
- **Protein B:** UniProt (6) accession of the interactor protein B
- **Naming Convention B:** naming convention for the interactor protein B. Only Swiss-Prot entries were used, so the value is "UniProtKB/Swiss-Prot/P" for every entry.
- **Synonyms B:** list of the protein name synonyms for the interactor protein B
- **Taxonomy ID B:** the taxonomy ID of the interactor protein B, since all the protein are human, the ID is 9606 for every entry.
- **Interaction Score:** the interaction score is defined as described in Ref. 1 and gives the probability of the interaction in a compartment dependent manner (Note that the interaction score is 0, if there was no localization information for one, or both of the interactors.)
- **Interaction Experimental System Type:** list of the experimental system types for the given interaction
- **Interaction Source Database:** list of the source databases for the given interaction as defined in the ComPPI (1) database.
- **Interaction PubMed ID:** list of the PubMed IDs for the sources of the given interaction.

# Supplementary References